\documentclass[a4paper,11pt]{article}
\pdfoutput=1

\usepackage{jheppub}

\usepackage{slashed}
\usepackage{xfrac}
\usepackage[table]{xcolor}
\usepackage{epstopdf}

\newcommand{\rr}[1]{\mathrm{#1}}
\newcommand{\cc}[1]{\mathcal{#1}}

\newcommand{\ff}[1]{\slashed{#1}}

\newcommand{\be}{\begin{eqnarray}}
\newcommand{\ee}{\end{eqnarray}}
\newcommand{\nn}{\nonumber\\}
\newcommand{\ba}{\begin{array}}
\newcommand{\ea}{\end{array}}
\newcommand{\bee}{\begin{equation}\ba{c}}
\newcommand{\eee}{\ea\end{equation}}

\newcommand{\bi}{\begin{itemize}}
\newcommand{\ei}{\end{itemize}}

\newcommand{\f}[2]{\frac{#1}{#2}}
\newcommand{\nf}[2]{\sfrac{#1}{#2}}

\newcommand{\co}{\mbox{c.c.}}

\title{\boldmath Limits on Vectorlike Leptons from Searches for Anomalous Production of Multi-Lepton Events}

\author{Radovan Derm\'i\v{s}ek,}
\author{Jonathan P. Hall,}
\author{Enrico Lunghi,}
\author{and Seodong Shin}

\affiliation{Physics Department, Indiana University,
                727 E.~Third St., Bloomington, IN 47405, USA}

\emailAdd{dermisek@indiana.edu}
\emailAdd{halljp@indiana.edu}
\emailAdd{elunghi@indiana.edu}
\emailAdd{shinseod@indiana.edu}

\abstract{
We consider extensions of the Standard Model by vectorlike leptons and set limits on a new charged lepton, $e_4^\pm$, using  the ATLAS search for anomalous production of multi-lepton events. It is assumed that only one Standard Model lepton, namely the muon, dominantly mixes with  vectorlike leptons resulting in possible decays $e_4^\pm \to W^\pm \nu_\mu$, $e_4^\pm \to Z\mu^\pm$, and $e_4^\pm \to h \mu^\pm$. We derive generally applicable limits on the new lepton  treating the branching ratios for these processes as free variables. We further interpret the general limits in two scenarios with $e_4^\pm$ originating predominantly from either the $SU(2)$ doublet or the $SU(2)$ singlet. The doublet case is more constrained as a result of larger production cross-section and extra production processes $e_4^\pm \nu_4$ and $\nu_4\nu_4$ in addition to $e_4^+ e_4^-$, where $\nu_4$ is a new neutral state accompanying $e_4$.  We find that  some combinations of branching ratios are poorly constrained, whereas some are constrained up to masses of more than 500~GeV. In the doublet case, assuming $\rr{BR}(\nu_4\rightarrow W\mu) = 1$, all masses below about 300~GeV are ruled out. Even if this condition is relaxed and additional decay modes, $\nu_4 \to Z \nu_\mu$ and $\nu_4 \to h \nu_\mu$, are allowed, below the Higgs threshold still almost all of the parameter space (of independent branching ratios) is ruled out. Nevertheless, even assuming the maximal production cross-section, which coincides with the doublet case, the new charged lepton can still be as light as the LEP-II limit allows. We discuss several possible improvements of current experimental analyses that would dramatically reduce the allowed parameter space, even with current data.
}

\begin{document} 
\maketitle
\flushbottom

\section{Introduction}

Among simplest extensions of the Standard Model (SM) are those with extra vectorlike quarks and leptons near the electroweak (EW) scale. The reason for SM fermions coming in three copies remains a mystery, and although having more than three families would not complicate the structure of the SM, extra chiral families are  ruled out experimentally. However, adding vectorlike partners along with the fourth family completely changes the picture. Vectorlike fermions can acquire masses independently of their Yukawa couplings to the Higgs boson and thus they are much less constrained. 

The possibility to contribute to basically any process in the SM in a controlled way (through Yukawa couplings) makes vectorlike fermions a simple framework for explaining various anomalies. Examples include attempts to explain the anomaly in the forward-backward asymmetry of the b-quark~\cite{Choudhury:2001hs, Dermisek:2011xu, Dermisek:2012qx} and the muon $g-2$ anomaly~\cite{Czarnecki:2001pv, Kannike:2011ng, Dermisek:2013gta}. Recently, their effects on Higgs boson decays were also studied. For example, they allow for a sizable modification of $h \to \gamma \gamma$~\cite{htogam} or they can significantly contribute to $h\to 4\ell$ or $2\ell 2\nu$, through flavor violating Higgs decays, thus possibly affecting measurements of $h \to ZZ^* $ and $h \to WW^*$~\cite{Falkowski:2014ffa, Dermisek:2014cia}. Even without couplings to SM fermions they were studied in connection with the unification of gauge couplings in the minimal supersymmetric model extended by a complete vectorlike family~\cite{Babu:1996zv} and the SM extended by several vectorlike families~\cite{Dermisek:2012as, Dermisek:2012ke}. For many other phenomenological implications of vectorlike fermions, see also ref.~\cite{Ellis:2014dza} and references therein.

In this paper, we use the ATLAS search for anomalous production of multi-lepton events~\cite{TheATLAScollaboration:2013cia} to set limits on a new lepton, $e_4^\pm$ (the lightest charged mass eigenstate originating from vectorlike pairs).\footnote{After the submission of this paper a new ATLAS analysis was released using different event categories and a different algorithm for identifying hadronic taus \cite{Aad:2014hja}.} Vectorlike leptons can be pair produced just like the SM leptons.  If they mix with the SM leptons they decay into a SM lepton plus $W$, $Z$, or Higgs boson $h$. We will focus on the scenario where only one SM lepton, namely the muon, dominantly mixes with the vectorlike leptons resulting in possible decays $e_4^\pm \to W^\pm \nu_\mu$, $e_4^\pm \to Z\mu^\pm$, and $e_4^\pm \to h \mu^\pm$. For the general set of couplings mixing the muon with vectorlike leptons, arbitrary values of branching ratios of $e_4^\pm$ into the three channels above can occur and are allowed by precision EW data. Hence we treat these branching ratios as free variables (constrained to sum to unity) in our analysis of the implications of the generally applicable limits that we derive. The limits in the case of mixing between the electron and vectorlike leptons are similar to those presented here. In the case of tau mixing the limits we obtain from the ATLAS data are about an order of magnitude weaker than in the muon case and do not constrain our parameter space with the exception of a small window near $105$ GeV for maximal branching ratios (BR($e_4 \to Z \tau$) = BR($\nu_4 \to W \tau$) = 1). A study of the impact of CMS data on vectorlike leptons decaying to taus is presented in Ref.~\cite{Falkowski:2013jya} where bounds somewhat stronger than ours are obtained. 

We consider two limiting scenarios with $e_4^\pm$ originating predominantly from either the $SU(2)$ doublet or the $SU(2)$ singlet. The predicted production cross-sections and thus the implied limits on branching ratios for these cases are different. However a more dramatic difference comes from the fact that the doublet $e_4^\pm$ is accompanied by a new neutrino $\nu_4$ and thus, in addition to $e_4^+ e_4^-$ production, also $e_4^\pm \nu_4$ and $\nu_4\nu_4$ production processes must be considered. If there are no new $SU(2)$ singlet neutrinos near the electroweak scale then the new neutrino, $\nu_4$, decays into $W^\pm \mu^\mp$. However, if there is also an $SU(2)$ singlet vectorlike neutrino near the EW scale, flavour violating couplings between SM neutrinos and the heavy neutrino can be generated, and in addition to $\nu_4 \to W^\pm \mu^\mp$, also $\nu_4 \to Z \nu_\mu$ and $\nu_4 \to h \nu_\mu$ should be considered. In the doublet case we assume $e_4^\pm$ and $\nu_4$ to be degenerate in mass. We analyse limits both with and without the assumption that $\nu_4$ decays into $W^\pm \mu^\mp$ with branching ratio one.

As long as vectorlike leptons decay promptly to light leptons, the constraints we study in this paper depend exclusively on branching ratios of vectorlike leptons and are insensitive to the precise values of the actual couplings. Electroweak precision data constrain couplings between vectorlike leptons and muons at the $10^{-2}$ level (see the analysis presented in Ref.~\cite{Dermisek:2014cia}). The requirement of prompt decays of vectorlike leptons implies couplings larger than about $10^{-6}$ (for which the displacement length is smaller than 100 $\mu$m). 

As opposed to new coloured fermions, the direct observational bounds on vectorlike leptons are only from LEP-II, where the lower mass bound is as weak as about 105~GeV~\cite{lep2}. It is therefore important to investigate the multi-lepton decay signatures of their Drell-Yan pair production at the LHC as well as searching for indirect signals in other measurements. Previously, limits on heavy leptons have been placed in ref.~\cite{Falkowski:2013jya} using the CMS analysis~\cite{CMS:2013jfa} which uses about 19.5~fb$^{-1}$ at 8~TeV to look quite generally at 3+ lepton final states, looking for new phenomena. Ref.~\cite{Falkowski:2013jya} considers a specific scenario where the branching ratios of $e_4$ to $Z\ell$ and $W\nu$ are fixed and decays involving the Higgs boson ($h\ell$) are not considered. In our paper we use the ATLAS multi-lepton search~\cite{TheATLAScollaboration:2013cia} with 20.3~fb$^{-1}$ at 8~TeV which also looks quite generally at 3+ lepton final states, although the categorization of the events and cut variables employed are somewhat different. (Leptons are electrons, muons, and hadronic taus; leptonically decaying taus show up as electrons and muons.) They set various bounds after splitting the data into four categories and then further applying cuts requiring different amounts of transverse energy in the form of leptons, jets, or missing energy or just generally.

The reason we concentrate on the ATLAS analysis is that they provide single lepton (electron, electron-from-tau, muon, muon-from-tau) and single hadronic tau fiducial efficiencies as well as particle level selection requirements; in fact, the ATLAS paper shows that a lack of precise knowledge of these efficiencies could lead to order of magnitude uncertainties on the expected limits on multi-lepton production (see figure~25 of ref.~\cite{TheATLAScollaboration:2013cia}).\footnote{In practice we generate events at the shower level rather than the particle level. We believe that this should not affect the results.}

This paper is organized as follows. In section~\ref{sec:model}, we briefly review the model that mixes one SM lepton generation with a vectorlike pair of leptons. In section~\ref{sec:method} we discuss our strategy to obtain limits on the new lepton states from the ATLAS search for anomalous production of multi-lepton events and provide the details of the analysis. The obtained limits are presented in section~\ref{sec:results} along with an analysis of the implications of the (generally applicable) limits in various scenarios. We discuss the results and conclude in section~\ref{sec:conclusions}.

\section{Model and Strategy}
\label{sec:model}
In this paper we consider the framework where the SM is simply extended by vectorlike pairs of new leptons $L_{L,R}$ and $E_{L,R}$. $E_R$ and $L_L$ have the same quantum numbers as the SM right-handed muon and left-handed muon doublet respectively and $E_L$ and $L_{R}$ are their vectorlike partners. We assume that these new fermions mix with the second generation of leptons. For simplicity we will not consider the case that new leptons mix with more than one SM lepton generation, in which case the limits from a variety of lepton flavour violating processes are expected to be stronger than direct production constraints. The most general Yukawa and mass terms for the muon and new leptons are
\be
& - \bar \mu_{L} y_{\mu} \mu_{R} H - \bar \mu_{L} \lambda^E E_{R} H   - \bar L_{L} \lambda^L \mu_{R} H -  \lambda \bar L_{L}  E_{R} H - \bar \lambda H^\dagger \bar E_{L}  L_{R}  & \nn
& - M_L \bar L_L L_R - M_E \bar E_L E_R + \co, &
\label{eq:lagrangian}
\ee
where the first term is the usual SM Yukawa coupling and is followed by Yukawa couplings between the muon and vectorlike leptons, those between vectorlike leptons, and direct mass terms for vectorlike leptons. After the EW symmetry is spontaneously broken the new extra leptons mix with the muon (and each other) to give new mass eigenstates $e_4$ and $e_5$ (in mass order), and flavour violating couplings  $e_4 - Z - \mu$, $e_4 - W - \nu_\mu$ and $e_4 - h - \mu$ are generated (see ref.~\cite{Dermisek:2014cia}).

\begin{figure}[tbp]
\centering
\includegraphics[width=.6\textwidth]{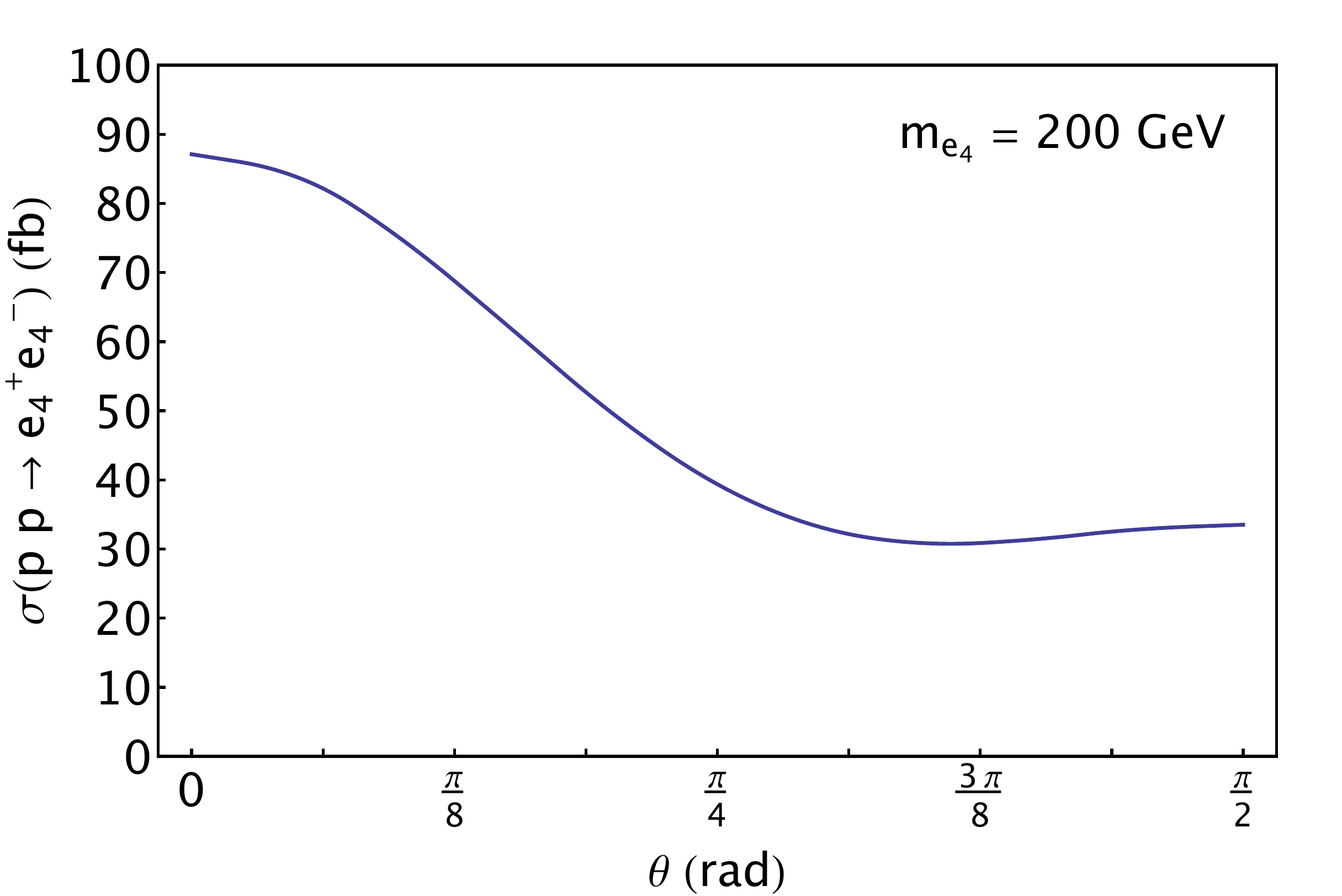}
\caption{Dependence of the $e_4^+ e_4^-$ production cross-section on the doublet-singlet mixing angle $\theta$ for $m_{e_4}=200\; {\rm GeV}$. Fully doublet and singlet leptons correspond to $\theta=0$ and $\theta=\pi/2$, respectively. For heavy vectorlike leptons, the shape of the curve is mostly independent of the mass. 
\label{fig:mixing}}
\end{figure}

In general the new charged states $L^\pm$ and $E^\pm$ mix and their production cross-section depends on the mixing angle (see figure~\ref{fig:mixing}). We illustrate the impact of the limits in the following two cases: the doublet case, where $e_4$ originates from the doublet $L$ (in this case we set $m_{e_4}= m_{\nu_4}$) and the singlet case, where $e_4$ originates from the singlet $E$. Limits on mixed scenarios can be inferred from these limiting cases with the aid of figure~\ref{fig:mixing}. We will then place limits on quantities of the form $\sigma(pp\rightarrow ab)\rr{BR}(a\rightarrow vm)\rr{BR}(b\rightarrow xn)$, where $a$ and $b$ are $e_4$ or $\nu_4$ with equal masses, $v$ and $x$ stand for massive vector bosons or SM Higgs boson, and $m$ and $n$ stand for muons or neutrinos (e.g. $\sigma(pp\rightarrow e_4^+ e_4^-)\rr{BR}(e_4^+\rightarrow W^+ \nu_\mu)\rr{BR}(e_4^-\rightarrow Z \mu^-)$).

It can easily be seen that the Lagrangian with the above terms added conserves a generalized muon number possessed by both the SM muon generation and the extra fields. The pair production of heavy states consists of one heavy state with generalised muon number $-1$ and another state with generalized muon number $+1$. In particular it is to be understood that in the doublet case $\sigma(e_4^\pm\nu_4)$ implies the production cross-section of $e_4^+$ with $\nu_4$ plus that of $e_4^-$ with $\bar{\nu}_4$ and $\sigma(\nu_4\nu_4)$ implies the production cross-section of $\nu_4$ with $\bar{\nu}_4$, where $\nu_4$ ($\bar{\nu}_4$) carries generalized muon number $+1$ ($-1$), and $\rr{BR}(\nu_4\rightarrow W\mu)$ implies $\rr{BR}(\nu_4\rightarrow W^+\mu^-) = \rr{BR}(\bar{\nu}_4\rightarrow W^-\mu^+)$.

When combining results there will be six interesting cases. First of all there will be the singlet case where the mass is below the threshold for decays of the heavy charged lepton into $h\mu$. In this case $\rr{BR}(e_4\rightarrow W\nu) = 1 - \rr{BR}(e_4\rightarrow Z\mu)$ and $\rr{BR}(e_4\rightarrow Z\mu)$ (say) is the only free variable. Above the Higgs threshold $\rr{BR}(e_4\rightarrow W\nu)$ (say) also becomes a free variable, but then $\rr{BR}(e_4\rightarrow h\mu)$ can be expressed as $1 - \rr{BR}(e_4\rightarrow Z\mu) - \rr{BR}(e_4\rightarrow W\nu)$. We proceed similarly for the doublet cases, but further split these two cases into four. With the particle content of the Lagrangian~(\ref{eq:lagrangian}) the $\nu_4$ decays exclusively to $W\mu$ ($\rr{BR}(\nu_4\rightarrow W\mu) = 1$). With the addition of new particles, e.g. extra vectorlike singlet neutrinos, the decay modes $\nu_4 \rightarrow h \nu$ and $\nu_4 \to Z\nu$ can be generated once all possible extra interactions are added to the Lagrangian~(\ref{eq:lagrangian}). Which branching ratios we will choose as the independent free variables in each case is summarised in table~\ref{tab:ver}.

\begin{table}[tbp]
\centering
\footnotesize
\begin{tabular}{|ll|ll|}
\hline
singlet, & below Higgs threshold &
$\rr{BR}(e_4\rightarrow Z\mu)$ &\\&&&\\\hline
singlet, & above Higgs threshold &
$\rr{BR}(e_4\rightarrow Z\mu)$, &
$\rr{BR}(e_4\rightarrow W\nu)$ \\&&&\\\hline
doublet, $\rr{BR}(\nu_4\rightarrow W\mu) = 1$, & below Higgs threshold &
$\rr{BR}(e_4\rightarrow Z\mu)$ &\\&&&\\\hline
doublet, $\rr{BR}(\nu_4\rightarrow W\mu) = 1$, & above Higgs threshold &
$\rr{BR}(e_4\rightarrow Z\mu)$, &
$\rr{BR}(e_4\rightarrow W\nu)$ \\&&&\\\hline
doublet, & below Higgs threshold &
$\rr{BR}(e_4\rightarrow Z\mu)$, &\\&&&
$\rr{BR}(\nu_4\rightarrow W\mu)$ \\\hline
doublet, & above Higgs threshold &
$\rr{BR}(e_4\rightarrow Z\mu)$, &
$\rr{BR}(e_4\rightarrow W\nu)$, \\&&
$\rr{BR}(\nu_4\rightarrow Z\nu)$, &
$\rr{BR}(\nu_4\rightarrow W\mu)$ \\\hline
\end{tabular}
\caption{The branching we choose as the independent free variables in the analysis combining results under different assumptions.
\label{tab:ver}}
\end{table}

\begin{figure}[tbp]
\centering
\includegraphics[width=.47\textwidth]{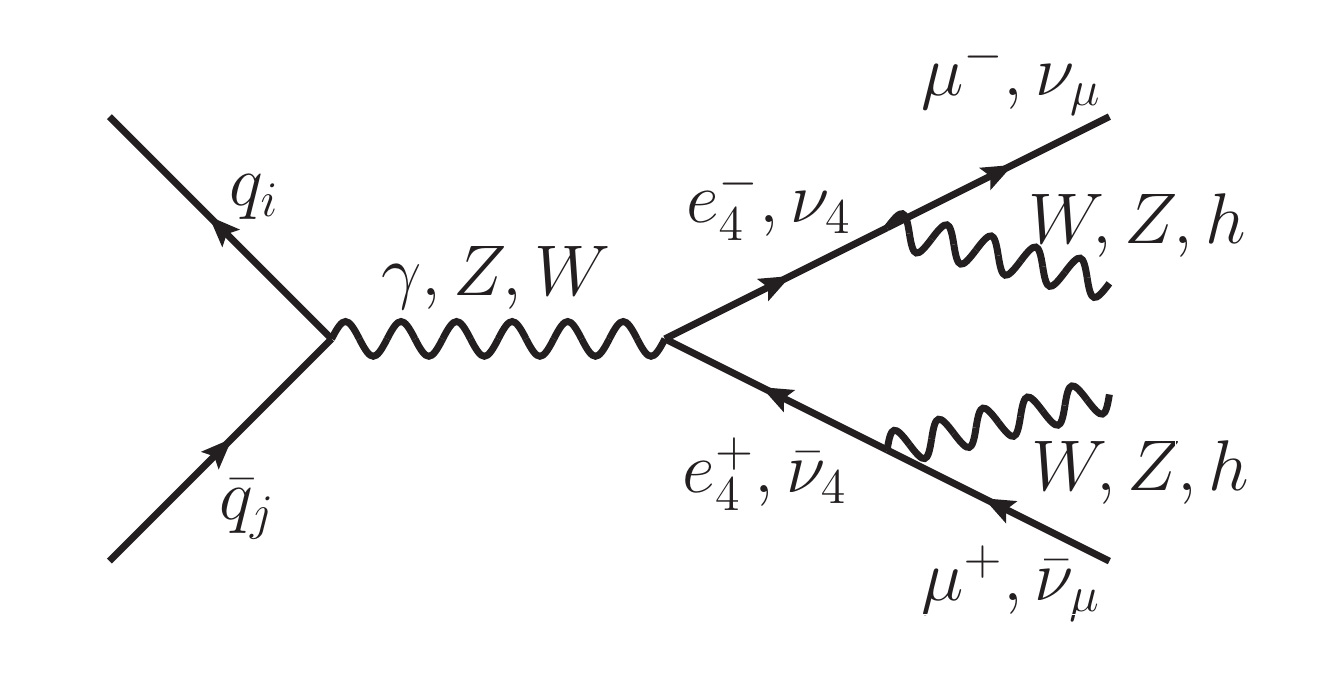}
\caption{Feynman diagrams for vectorlike charged and neutral leptons' production and decays. All combinations satisfying electric charge conservation are allowed, keeping in mind that the photon does not couple to heavy neutrinos.
\label{fig:VLL}}
\end{figure}

The strategy is to produce event samples of 20 possible processes ($pp\to e_4^+ e_4^-, e_4^\pm \nu_4, \nu_4 \nu_4$ with each heavy lepton decaying in $W$, $Z$, or $h$ plus light lepton), with the heavy vector and Higgs bosons decaying as per their Standard Model branching ratios. These processes are summarised in figure~\ref{fig:VLL}. We can then apply the various sets of cuts from the ATLAS analysis~\cite{TheATLAScollaboration:2013cia} and apply each relevant bound to see which bound is the most constraining for each sample process. As well as the cuts in the papers, we will also implement a simple five-lepton cut relevant for the ATLAS detector and triggers and compare the expected number of events with the zero expected background and assuming zero observed events. These five-lepton results will give an indication of where this kind of cut could be effective, but technically they do not apply since we do not have confirmation that the observed number of events for this cut is zero.

There is one quantity of the above form that cannot be constrained by the considered ATLAS analysis since the corresponding process cannot produce more than two leptons; this quantity is $\sigma(e_4^+e_4^-)\rr{BR}(e_4\rightarrow W\nu)^2$. The best limits on this quantity would seem to come from limits on chargino pair production where each chargino decays to a $W$ and a massless, stable neutralino. Such limits are placed in ref.~\cite{Aad:2014vma}, where they require two leptons and utilize $m_{T2}$. The inferred 95\% limits on $\sigma(e_4^+e_4^-)\rr{BR}(e_4\rightarrow W\nu)^2$ at $m_{e_4}=100$, 150, 200, and 250~GeV are roughly 115, 25, 10, and 10~pb respectively. These limits are not really constraining for typical sizes of heavy lepton production.

\section{Method}
\label{sec:method}

In this section we explain in detail how the events are generated and the analyses proceed. For convenience we divide this section into a description of our event generation, details of the analysis implementation, how we obtain the limits, and how we apply the additional five-lepton cut.

\subsection{Event generation}
\label{sec:eventgen}

The dominant process for pair production is expected to be Drell-Yan ($pp\to \gamma,Z \to e_4^+ e_4^-$, $pp\to Z \to \nu_4 \nu_4$). Production via an $s$-channel SM Higgs boson cannot be large unless it is resonant and this would require too light new leptons. In the doublet $e_4$ case production of $e_4^\pm\nu_4$ via a $W^\pm$ ($pp\to W^\pm \to e_4^\pm \nu_4$) should be considered alongside the Drell-Yan $e_4^+e_4^-$ and $\nu_4 \nu_4$ production. We find that the limits do not depend on the doublet/singlet nature of the $e_4^\pm$. Kinematically, the only differences between these processes are the different proportions of Drell-Yan production via $\gamma$ and $Z$ (pure singlet and doublet $e_4^\pm$ have identical couplings to photons but different couplings to the $Z$); we find that the kinematic distributions are approximately independent of this consideration. Therefore we will just quote limits on various products of production cross-sections and branching fractions for different masses. These limits will then be compared to different predicted production cross-sections to see if the relevant branching ratios are actually constrained.

For the cases involving only decays into massive vector bosons (not Higgs) the events are generated with {\tt SHERPA 2.0}~\cite{sherpa} at the parton level; vector bosons and tau leptons decays are handled by the {\tt HARD\_DECAYS} module of {\tt SHERPA}. For the cases including decays to Higgs bosons the events are generated with {\tt MadGraph5}~\cite{madgraph5} and the subsequent decays of the Higgs and of the vector bosons are handled by {\tt Pythia6}~\cite{Sjostrand:2006za}, which is also used to add initial and final state radiation; tau leptons produced from the vector or Higgs bosons are decayed with {\tt TAUOLA}~\cite{tauola} in the {\tt MadGraph5 pythia-pgs} package. The relevant flavor violating couplings between the heavy and light leptons have been implemented with {\tt FeynRules}~\cite{feynrules}. We confirm that we derive the same efficiencies using these two event generation schemes within statistical errors due to generating only finite numbers of events. We generate enough events in each case such that the best cut efficiency is calculated with sufficient accuracy (at least 100K events per final state).

The {\tt HepMC} and {\tt StdHEP} event files produced from the above Monte Carlo event generation schemes are converted into CERN {\tt root} files with a custom tree that mimics the {\tt HepMC} vertex structure and contains all of the information that we need for the analysis. The analysis is implemented as a {\tt root} macro, in which the jet clustering algorithm is implemented using {\tt FastJet}~\cite{fastjet}, and proceeds as explained in the following subsection.\footnote{The converter and analysis codes can be made available upon direct request to one of the authors.}

\subsection{Analysis}

First, each generated event is analysed and the final state particles are identified. Decaying (intermediate state) tau leptons are also identified. In this way muons and electrons coming from tau decays can be distinguished from those that do not come from tau decays. For each hadronically decaying tau lepton in the event a hadronic tau is defined that has the four-momentum of decaying tau lepton minus the four-momentum of the neutrino from the decay, i.e. the four-momentum of the visible tau decay products as defined in~\cite{TheATLAScollaboration:2013cia}. All final state particles except for muons (and neutrinos) are passed to the jet clustering algorithm. Jets are clustered using an anti-$k_t$ algorithm with distance parameter $\Delta R=0.4$ implemented using {\tt FastJet}. The neutrino four-momenta are summed to determine the missing momentum, $\ff{p}_T$, and missing transverse energy, $\ff{E}_T\equiv |\ff{p}_T|$, in the event. $p_T$ and $\eta$ cuts are then implemented on the reconstructed jets and on the leptons and hadronic taus.

Next, the fiducial efficiencies are taken care of. We split each event into all possibilities for each lepton and hadronic tau being kept or lost, using the tables of single lepton (and hadronic tau) fiducial efficiencies in the ATLAS paper~\cite{TheATLAScollaboration:2013cia}. This is where one needs to know whether the leptons are from tau decays or not. Next, some overlapping reconstructed particles are neglected as explained in the paper. The event (each possibility for each event) is then tested to make sure that there are either 3+ reconstructed electrons and muons or 2 reconstructed electrons and muons and 1+ reconstructed hadronic taus. The leading reconstructed electron or muon is then tested to make sure it has $p_T > 25$~GeV; this is the final event selection requirement.

The events (each possibility for each event) are then categorized into one of four mutually exclusive categories. If the event passed the event selection by having 3+ reconstructed electrons and muons then the three leading leptons are the ``leptons that define the event'' and the event is categorized as ``3$e/\mu$''. Alternatively, if the event needed an extra hadronic tau to pass the event selection then the three leptons that define the event are the electrons and muons and the leading hadronic tau and the event is categorized as ``2$e/\mu$+$\tau$''. The events are also categorized by whether or not there is a $Z$ candidate. $Z$ candidates are formed only out of the 3 leptons that define the event.\footnote{This is not explicit in the paper, but is shown in the {\tt Rivet}~\cite{Buckley:2010ar} implementation {\tt ATLAS\_2012\_I1204447} of the previous ATLAS analysis~\cite{Aad:2012xsa}, where the 7~TeV data is analysed, and was confirmed in a private communication with the main author of that {\tt Rivet} analysis.} An event is categorized as ``on-$Z$'' if there is an opposite-sign-same-flavour (OSSF) pair (or OSSF pair plus an electron) with invariant mass within 20~GeV (either side) of the $Z$ mass. The four event categories are then:
\bi
\item $\ge$3$e/\mu$ on-$Z$,
\item 2$e/\mu$+$\tau_h$ on-$Z$,
\item $\ge$3$e/\mu$ off-$Z$,
\item 2$e/\mu$+$\tau_h$ off-$Z$.
\ei
Since not all reconstructed leptons are necessarily used to form $Z$ candidates we find that a large proportions of events that do contain $Z$ bosons decaying to leptons can become classified as off-$Z$ if there are leptons from other sources (or more than one $Z$).

Each event is then tested to see whether it passes further cuts. These additional cuts are defined in terms of the following variables:\bi
\item $H_T^l$---the scalar sum of $p_T$s of the three leptons that define the event,
\item min $p_T^l$---the minimum $p_T$ of the three leptons,
\item $H_T^j$---the magnitude of the vector sum of all jet $p_T$s,
\item $m_{\rr{eff}}$---the scalar sum $\ff{E}_T+H_T^j+H_T^l$,
\item $\ff{E}_T$---the missing transverse energy.
\ei
They also present results for $b$-tag cuts and we will present representative limits from these cuts. Since exact fiducial $b$-tagging efficiencies are not given in the ATLAS analysis, to obtain representative limits we simply apply a flat 85\% (which is the working point used) $b$-tagging efficiency to reconstructed jets that contains $b$s. We find that this cut can be effective in the low mass region for processes involving at least one Higgs boson. Even though our $b$-tagging implementation is approximative, we decided to include these limits when they are stronger than those obtained from the other cuts. All the sets of cuts are summarised in table~1 of the ATLAS paper~\cite{TheATLAScollaboration:2013cia}.

\subsection{Extraction of the limits}
\label{ssec:calc}

We will label the quantities that we want to place limits on by the index $i$ and call them $\sigma_i$. The total number of events for a given cross-section $\sigma_i$ and set of cuts $k$ in category $j$ will be
\be
N_{ijk} &=& L\sigma_i\varepsilon_{ijk},
\ee
where $L$ is the integrated luminosity and $\varepsilon_{ijk}$ is the signal efficiency which also takes into account the fiducial efficiency. The limits given in table~14 onwards in the ATLAS paper (``observed'' column) are 95\% C.L. limits on the number of events surviving cuts per unit integrated luminosity, $N_{95}/L$, and are given in fb. We therefore have the set of conditions
\be
L\sigma_i\varepsilon_{ijk} &<& N_{95}\nn
\Rightarrow\quad \sigma_i &<& \f{N_{95}}{L}\f{1}{\varepsilon_{ijk}}
\ee
at 95\% C.L.. One categorisation $j$ and set of cuts $k$ will then set the best limit on each $\sigma_i$. To set limits all we need to calculate is the set of efficiencies $\varepsilon_{ijk}$.

\subsection{Five-lepton cut}

Before the categorization stage (but with the rest of the analysis remaining the same) we also implement a five-lepton cut that simply requires that there are five reconstructed electrons and muons.

Let $n$ be the observed number of events that we will set to zero and let $\mu$ be the expected number of events
\be
\mu &=& L(\sigma_b\epsilon_b+\sigma_i\epsilon_i),
\ee
where $\sigma_b\epsilon_b$ is the expected background cross-section times the background efficiency and $\sigma_i\epsilon_i$ is the cross-section we want to place a limit on times its efficiency for the five-lepton cut; $L$ is the integrated luminosity. To get sensible 95\% C.L. limits the question to ask is what is the cross-section $\sigma_i$ such that~\cite{Astone:2000kg}
\be
0.05 &=& \f{p(n|\sigma_i)}{p(n|\sigma_i=0)}.
\ee
Here the likelihoods $p$ are given by the Poisson distribution $\cc{P}$
\be
p(n|\sigma_i) &=& \cc{P}(n|\mu) = \f{\mu^n e^{-\mu}}{n!}.
\ee
Rearranging and setting $n=0$ yields
\be
0.05 = e^{-L\sigma_i\epsilon_i}
&\quad\Rightarrow\quad & \sigma_i = \f{-\ln(0.05)}{L\epsilon_i},
\ee
where $-\ln(0.05)=3.00$. Note that this limit is independent of the expected background $\sigma_b$ as long as $n=0$. All we need to know is the efficiency $\epsilon_i$ and the integrated luminosity $L$.

\section{Results}
\label{sec:results}

In this section we present our results. For convenience we divide this section into a summary of the generally applicable limits that we derive and analyses of the limits on branching ratios that these general limits imply under various assumptions.

\subsection{Generally applicable limits compared to predicted cross-sections}

\begin{figure}[tbp]
\centering
\includegraphics[width=.45\textwidth]{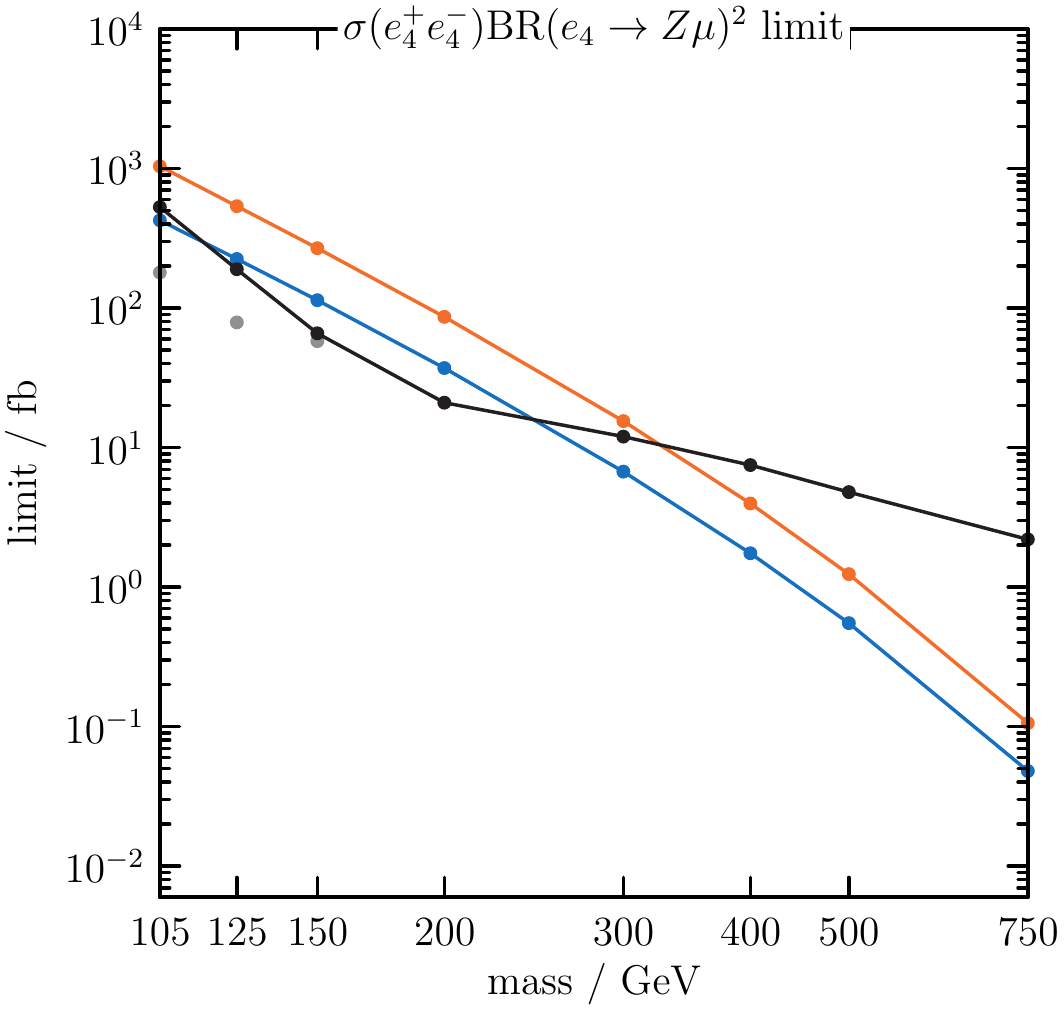}\hfill
\includegraphics[width=.45\textwidth]{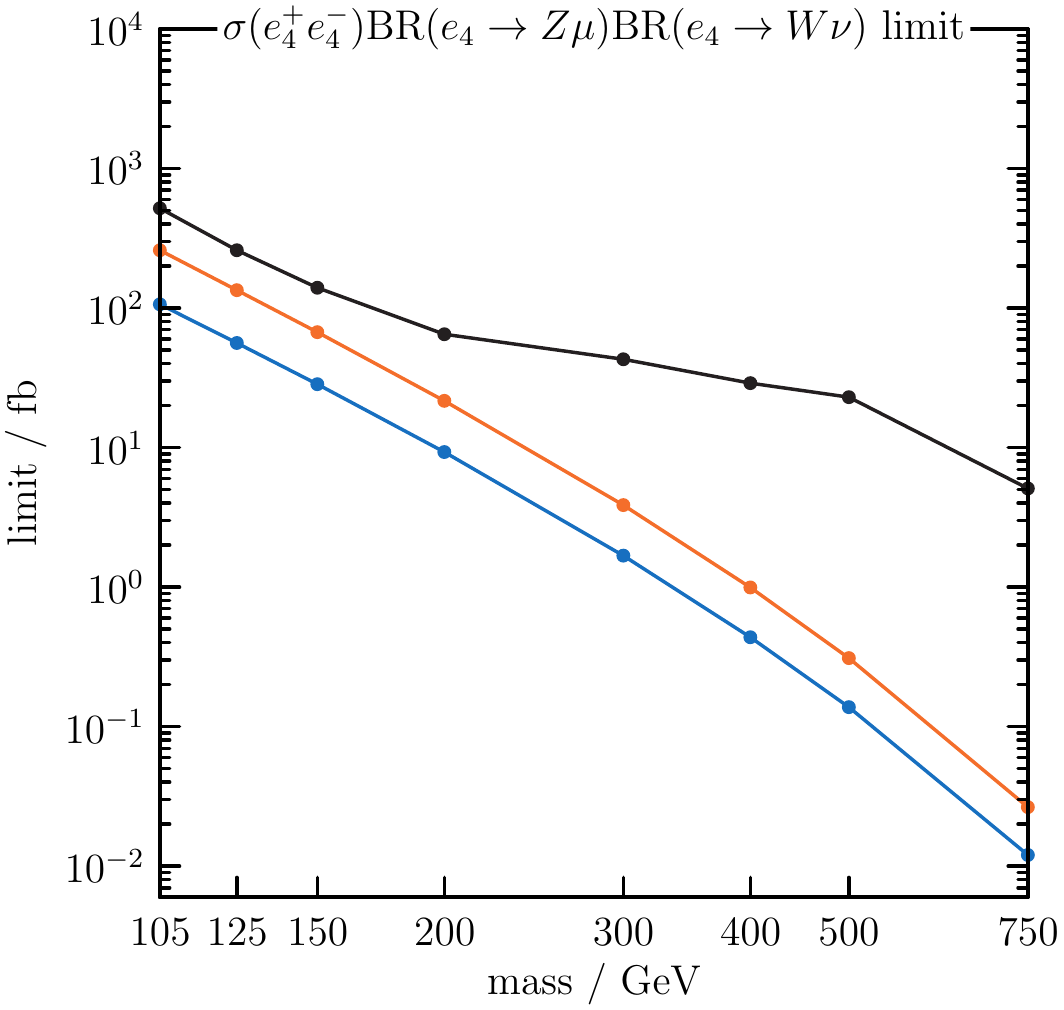}\\\hfill
\includegraphics[width=.45\textwidth]{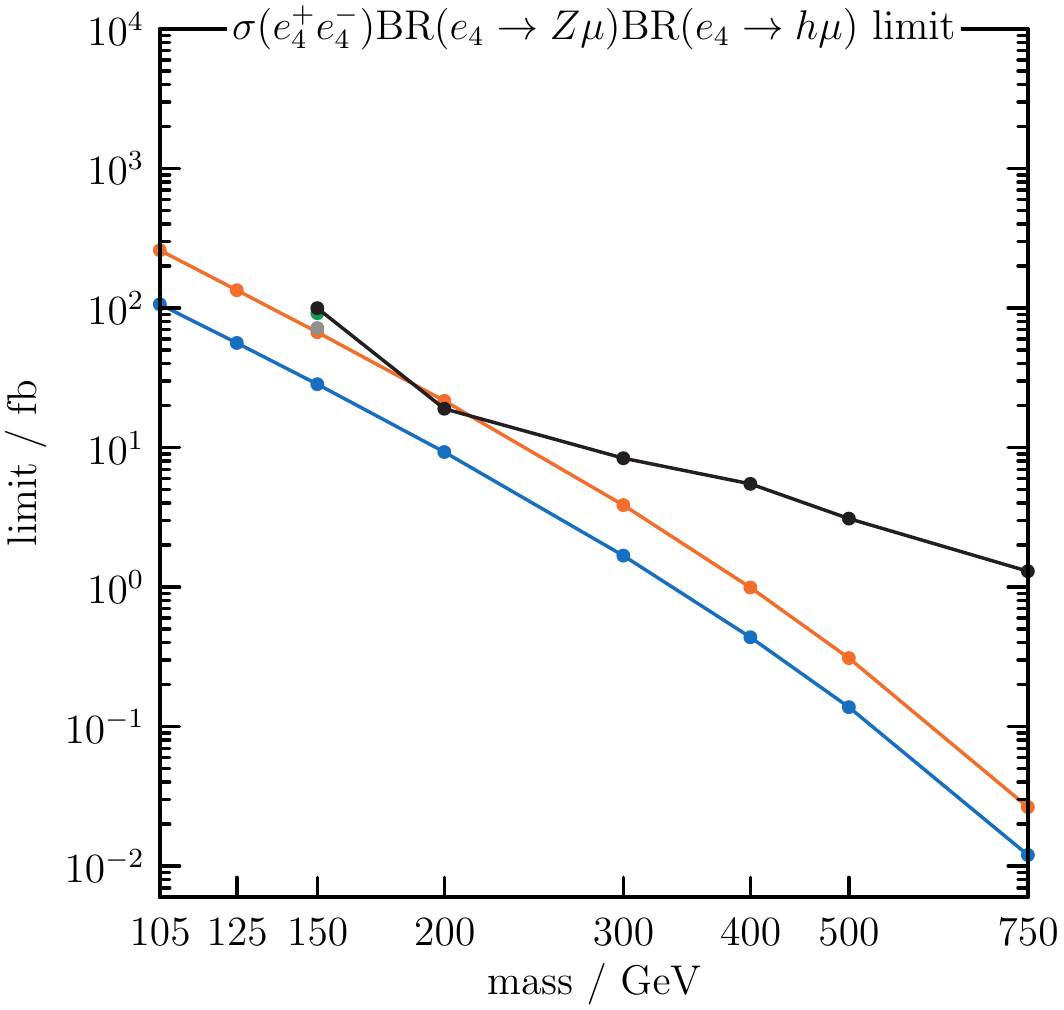}\\
\includegraphics[width=.45\textwidth]{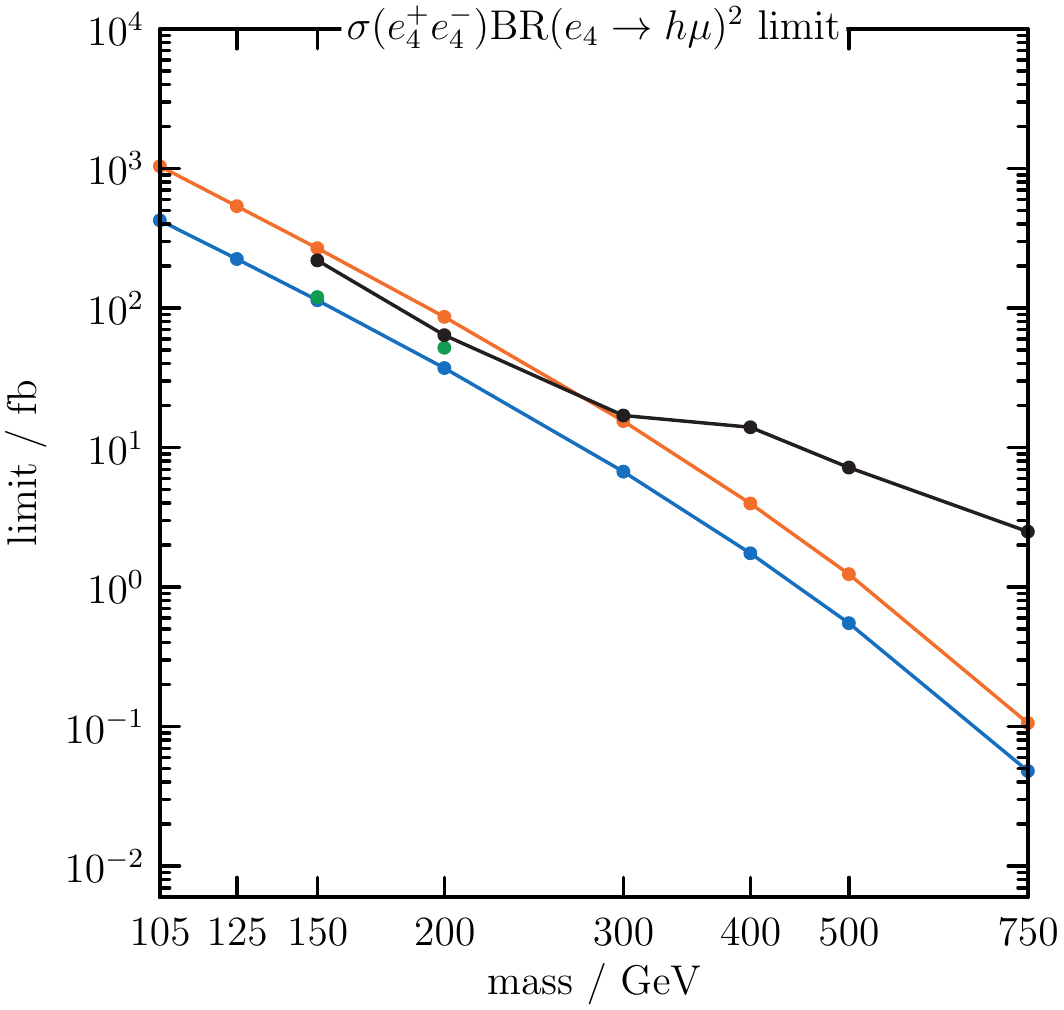}\hfill
\includegraphics[width=.45\textwidth]{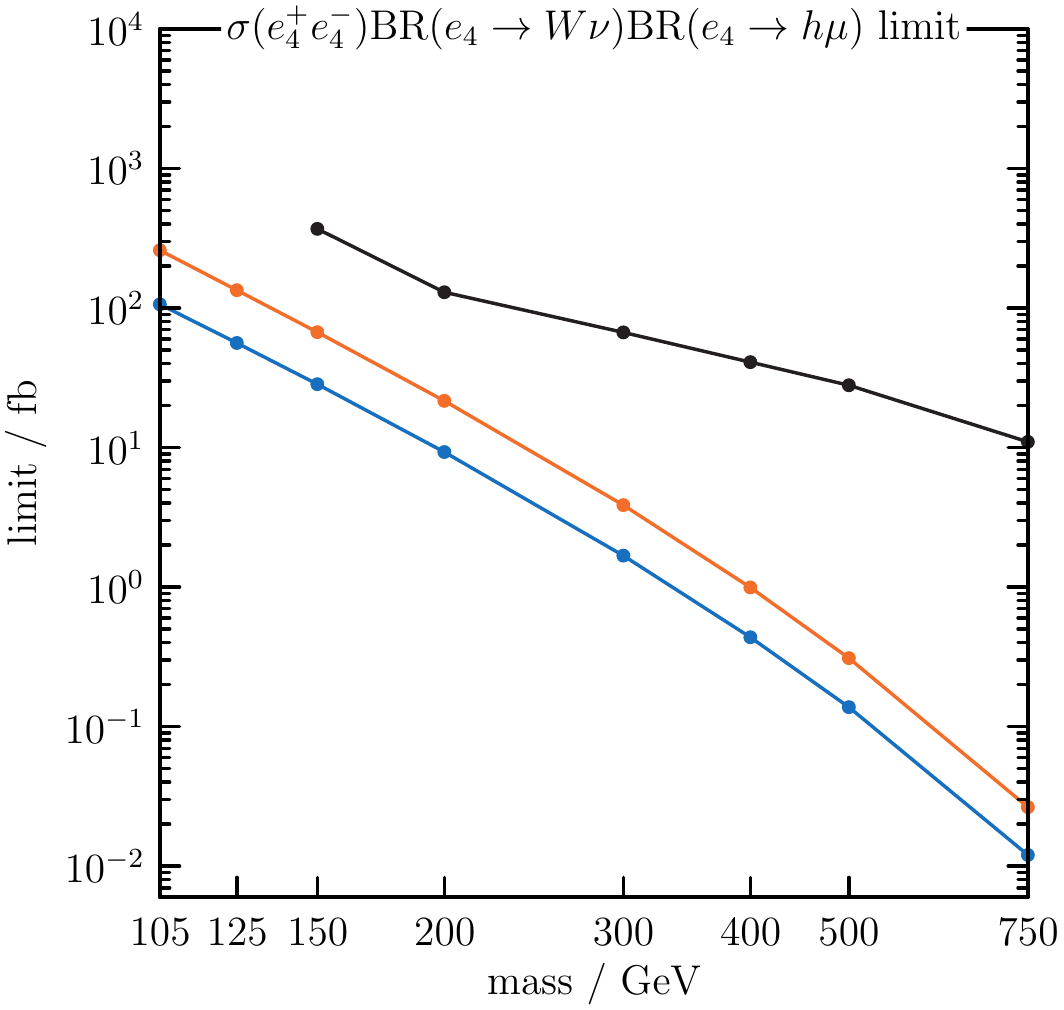}
\caption{The black points show limits on $\sigma(e_4^+ e_4^-)$ times a product of branching ratios for different masses. Grey points show the limits coming from the five-lepton cut in the cases where this provides a better limit (assuming zero data). Similarly, green points points are the representative limits from the $b$-tag cuts in the cases where these are better. In the left plots the orange and blue points are the predicted production cross-section $\sigma(e_4^+ e_4^-)$ in the doublet and singlet case respectively. In the right plots these production cross-sections are multiplied by $\nf{1}{4}$. Thus when a black point is below a coloured point a non-trivial limit is set on the relevant product of branching ratios assuming that production cross-section. $\sigma(e_4^+ e_4^-){\rm BR}(e_4 \to W\nu)^2$ is missing because it does not provide at least 3 charged leptons in the final state. 
\label{fig:EE}}
\end{figure}

\begin{figure}[tbp]
\centering
\includegraphics[width=.45\textwidth]{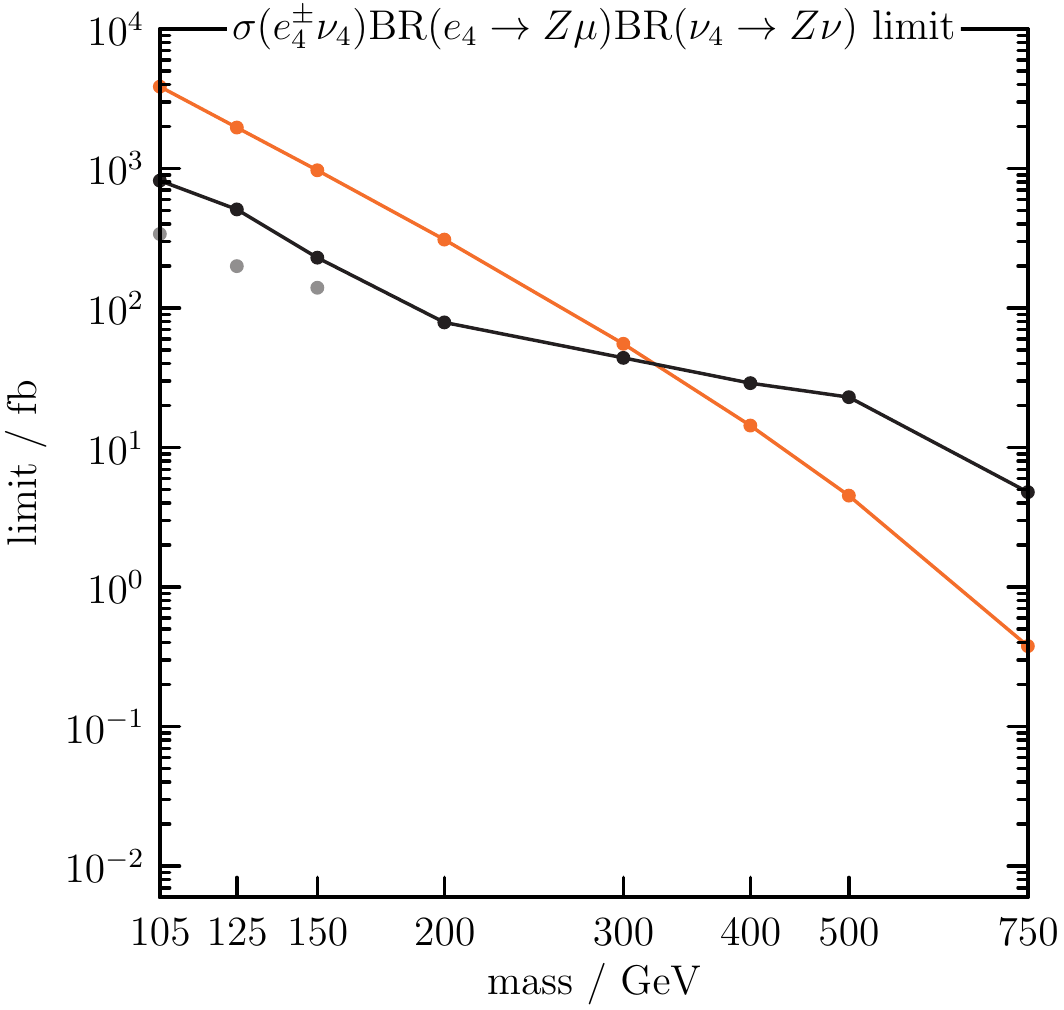}\hfill
\includegraphics[width=.45\textwidth]{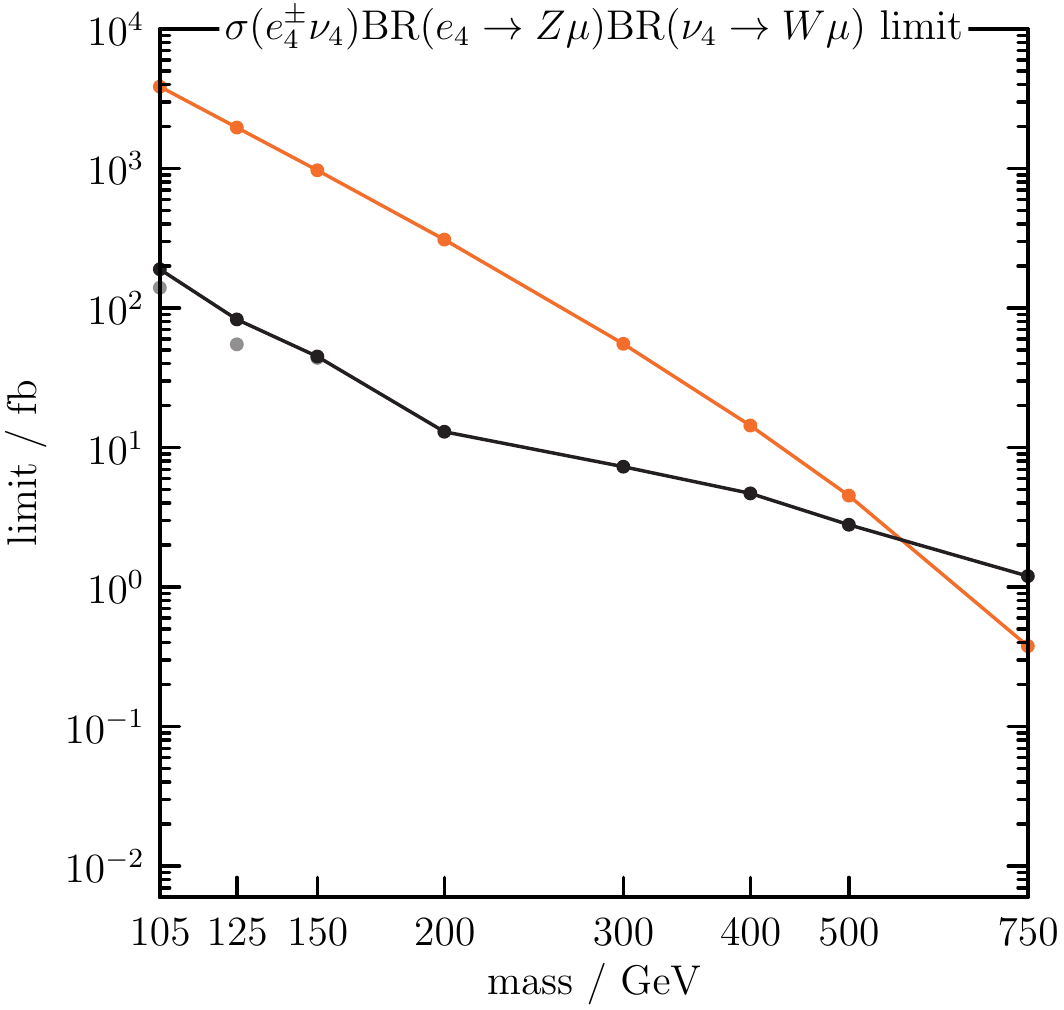}\\
\includegraphics[width=.45\textwidth]{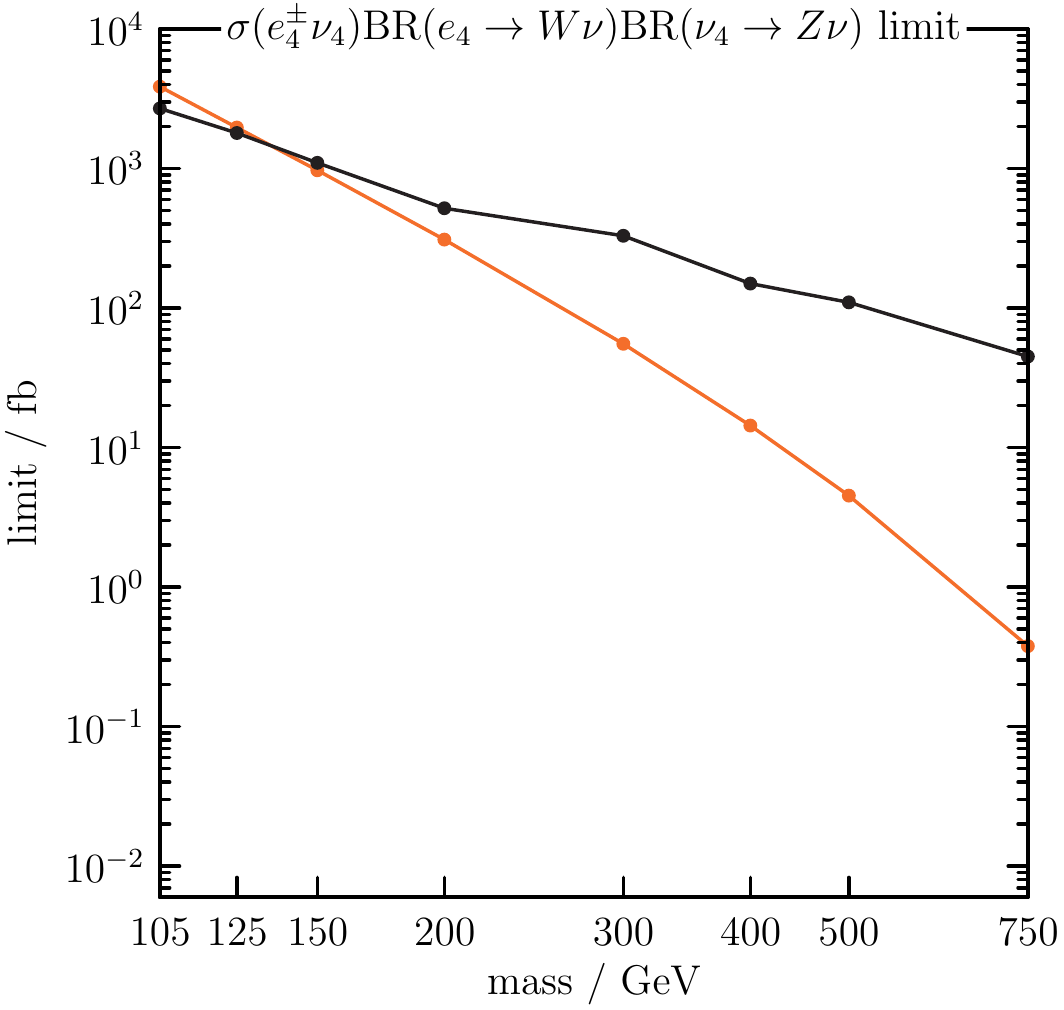}\hfill
\includegraphics[width=.45\textwidth]{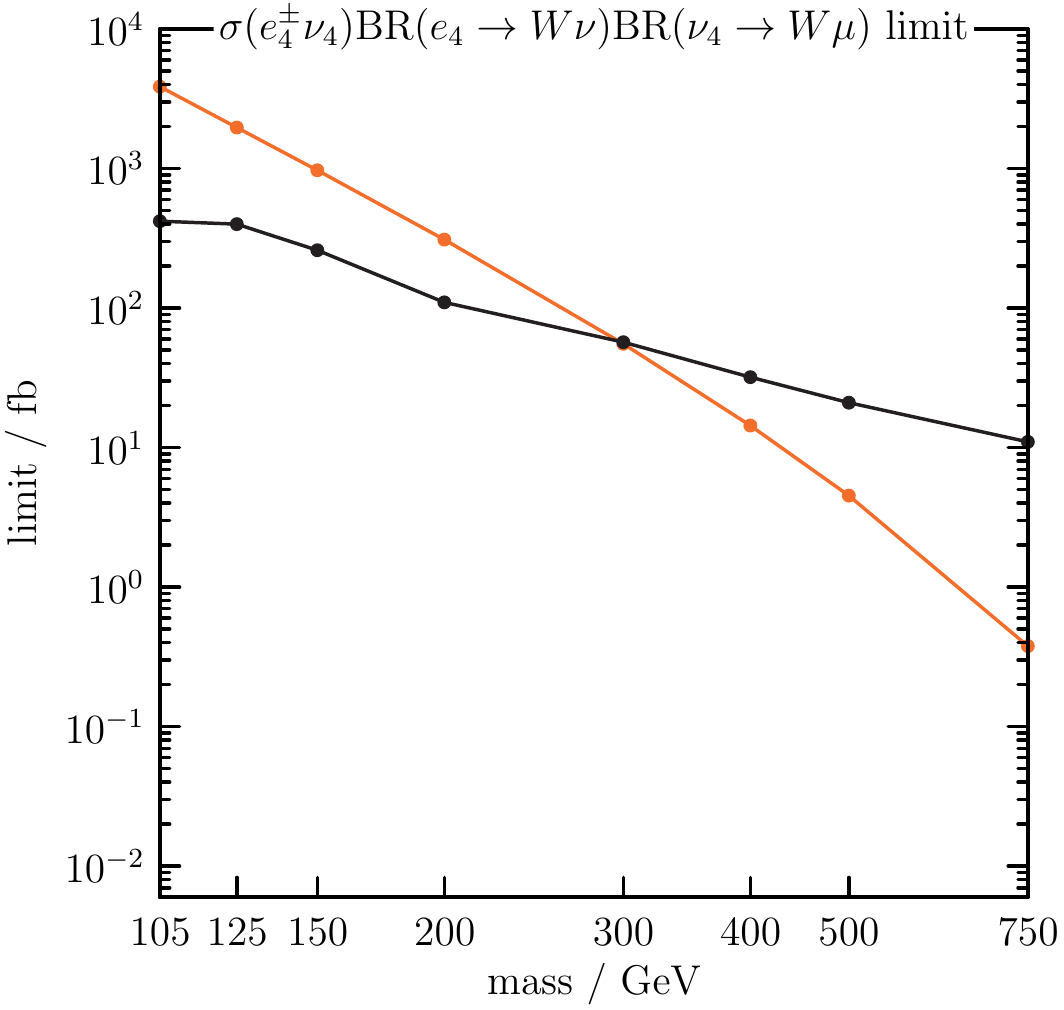}
\caption{The black points show limits on $\sigma(e_4^\pm \nu_4)$ times a product of branching ratios for different masses. Grey points show the limits coming from the five-lepton cut in the cases where this provides a better limit (assuming zero data). Similarly, green points points are the representative limits from the $b$-tag cuts in the cases where these are better. The orange points are the predicted production cross-section $\sigma(e_4^\pm \nu_4)$. Plots continue in figure~\ref{fig:ENb}.
\label{fig:ENa}}
\end{figure}

\begin{figure}[tbp]
\centering
\includegraphics[width=.45\textwidth]{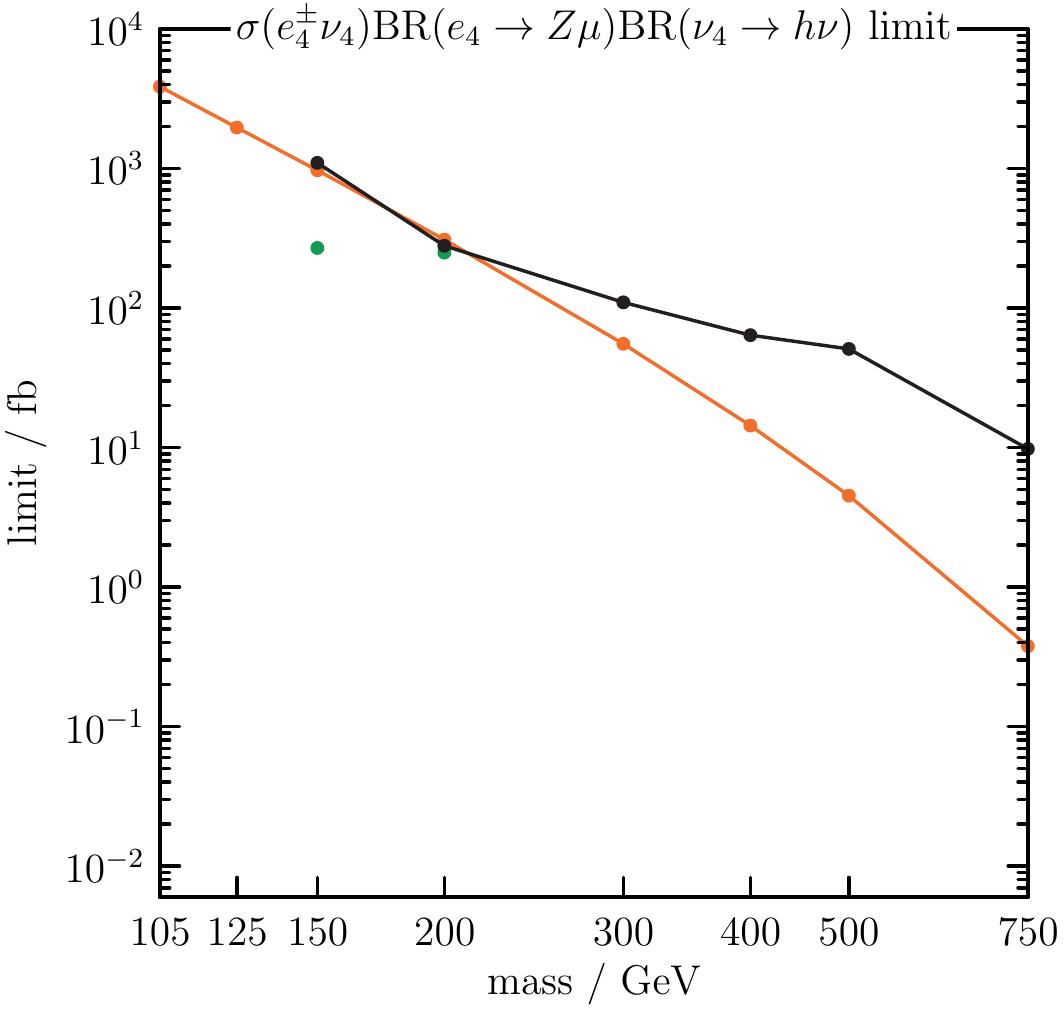}\hfill
\includegraphics[width=.45\textwidth]{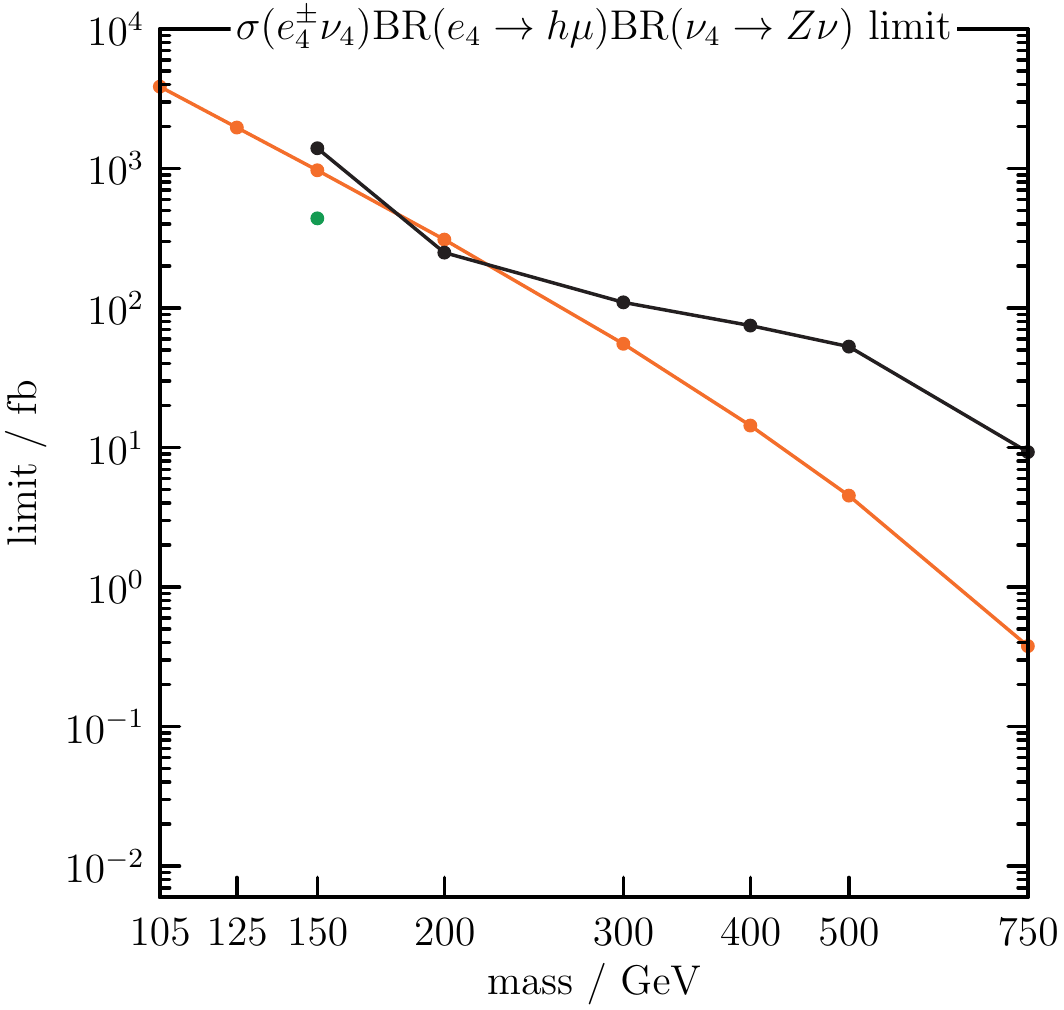}\\
\includegraphics[width=.45\textwidth]{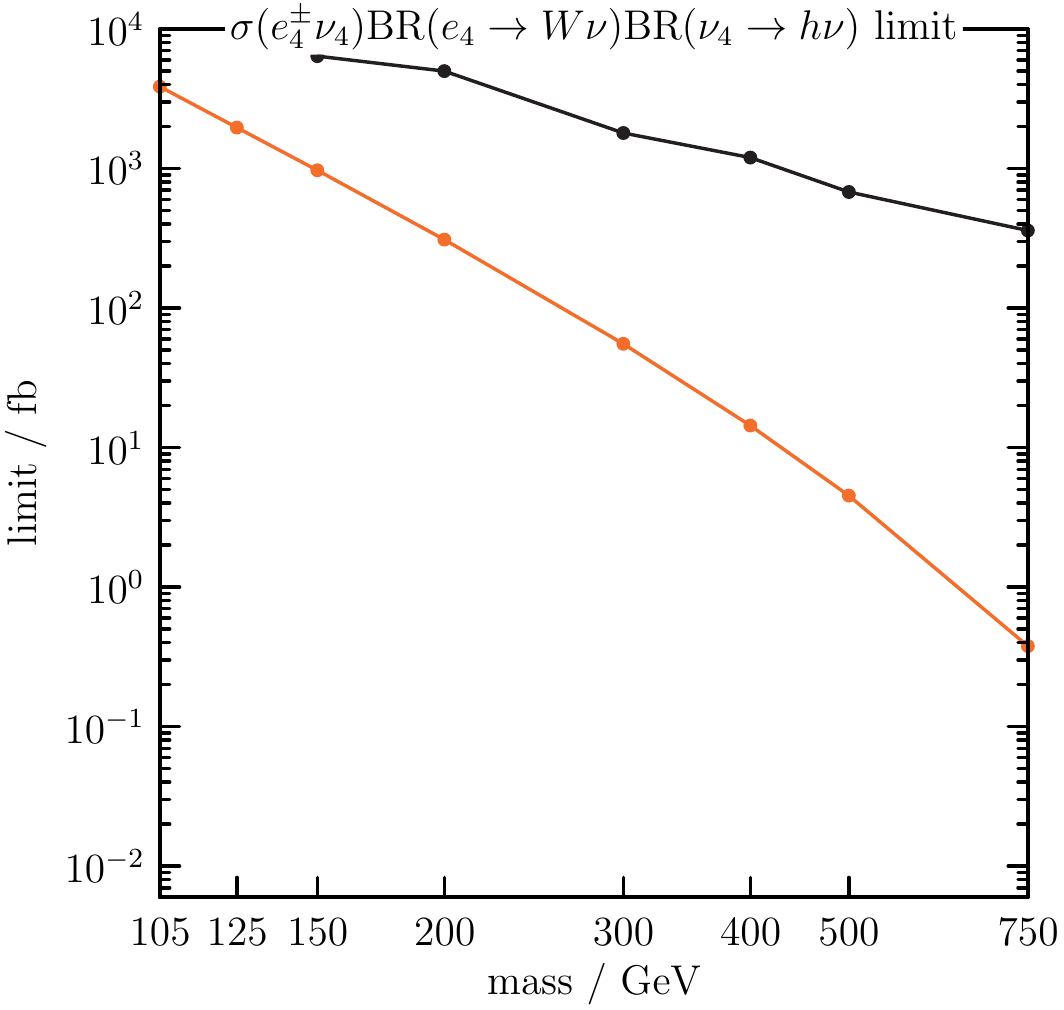}\hfill
\includegraphics[width=.45\textwidth]{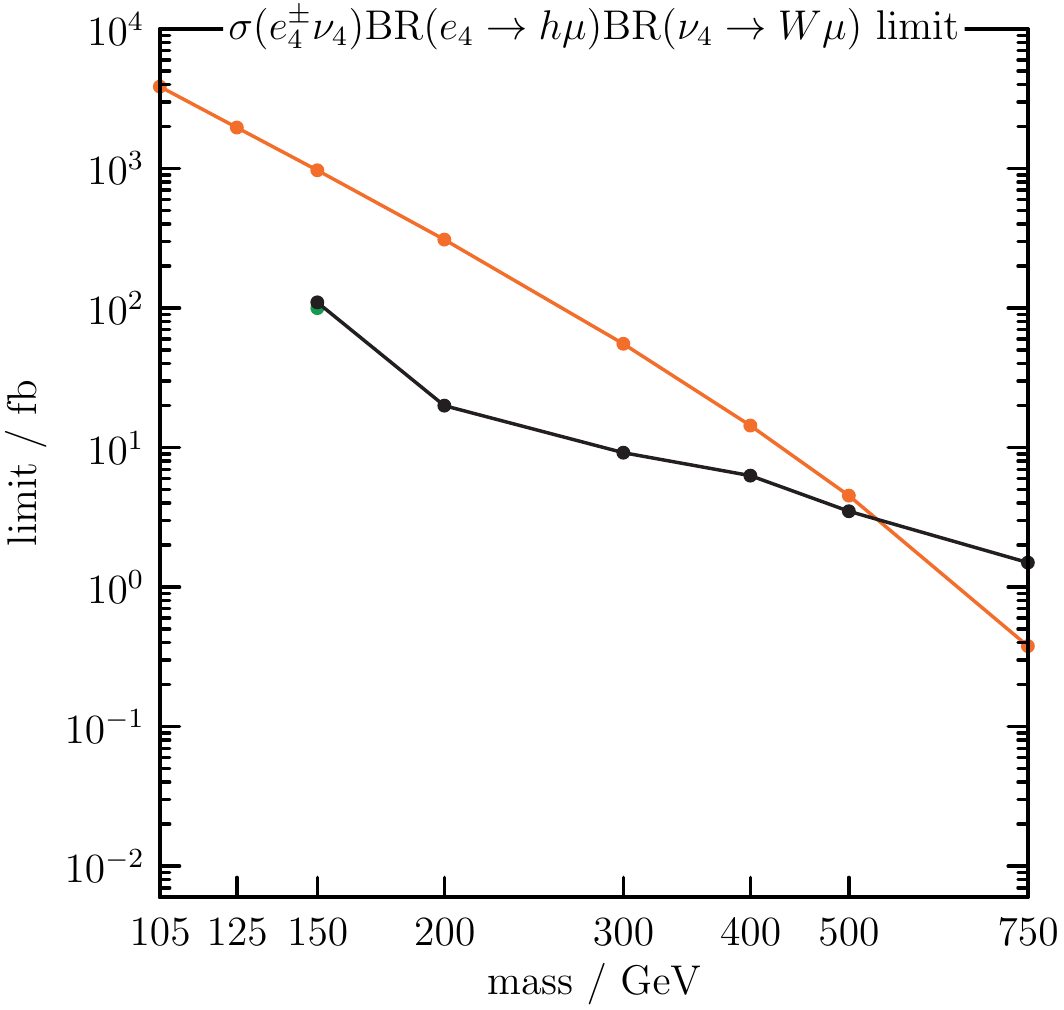}\\
\includegraphics[width=.45\textwidth]{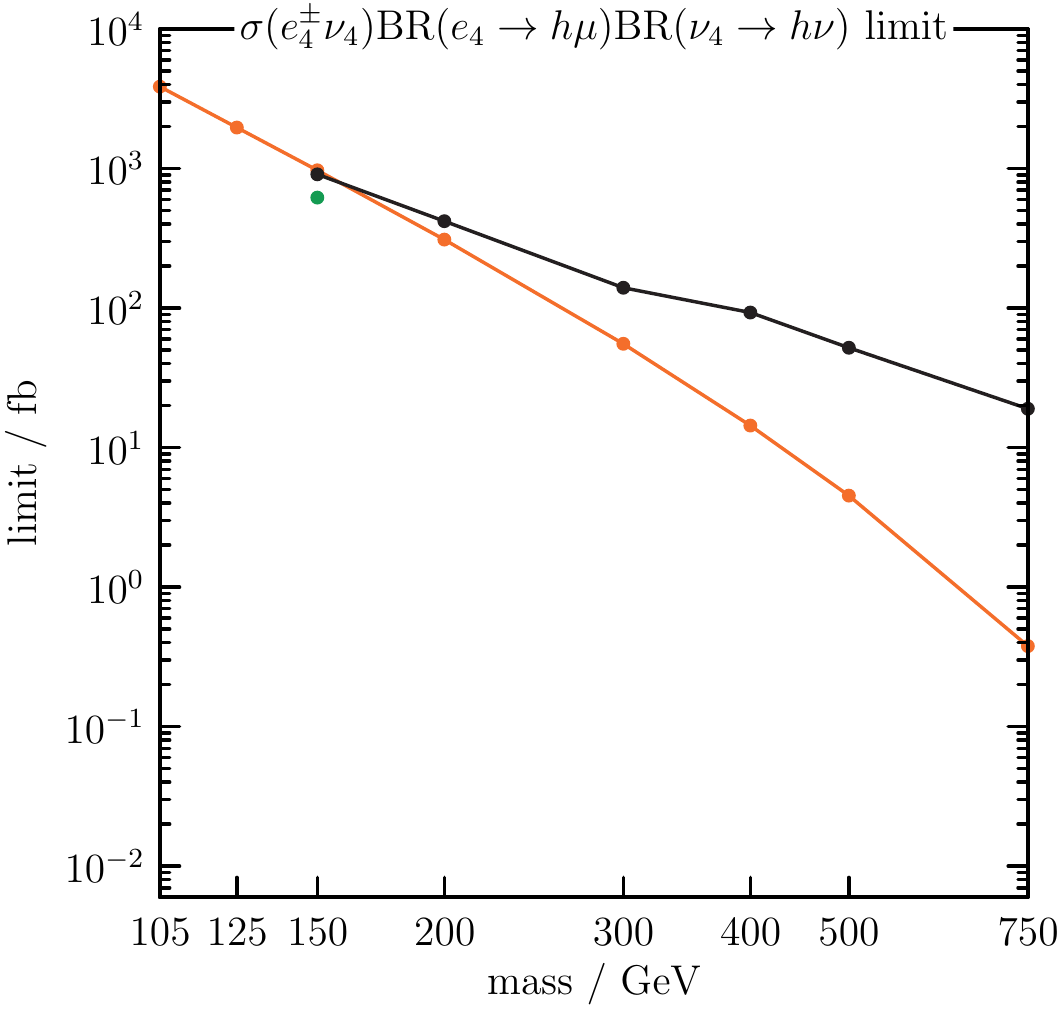}
\caption{Plots continuing from figure~\ref{fig:ENa}.
\label{fig:ENb}}
\end{figure}

\begin{figure}[tbp]
\centering
\includegraphics[width=.45\textwidth]{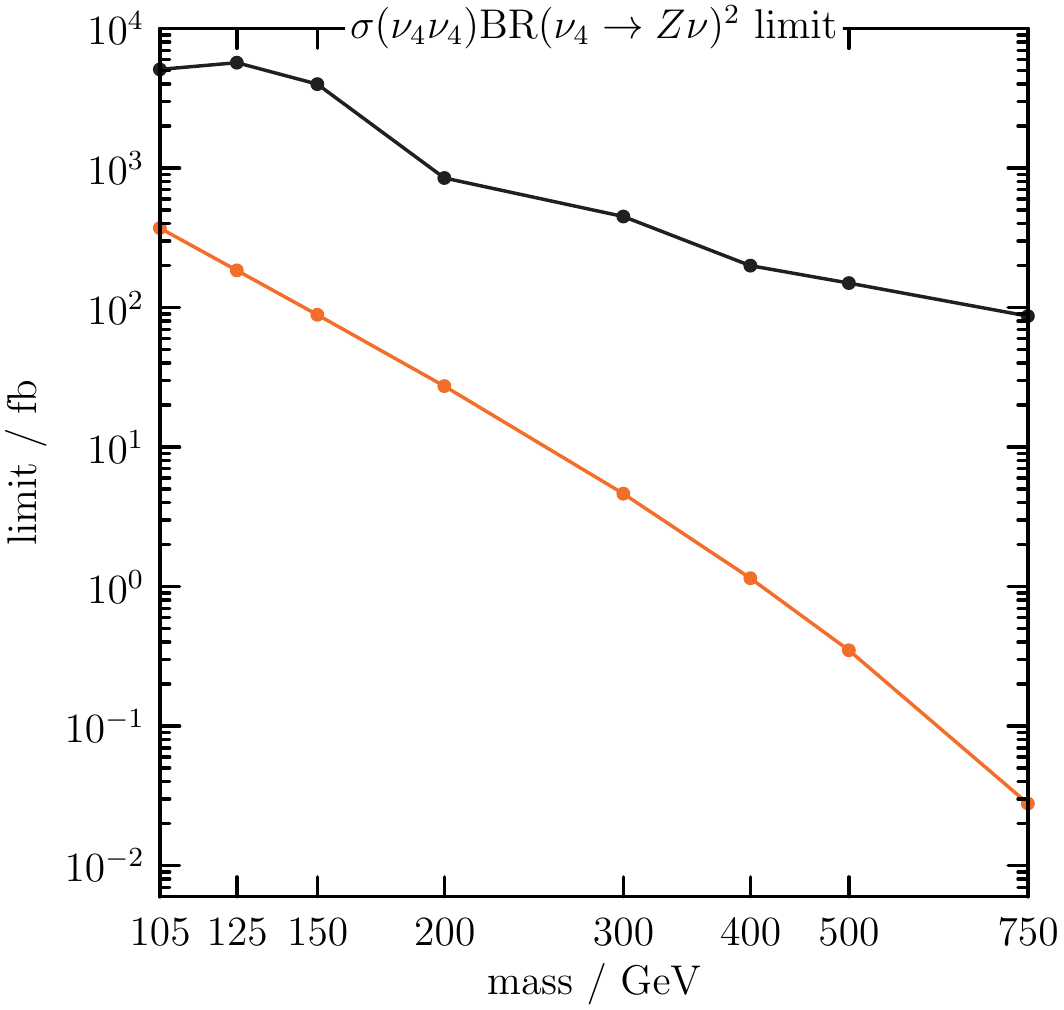}\hfill
\includegraphics[width=.45\textwidth]{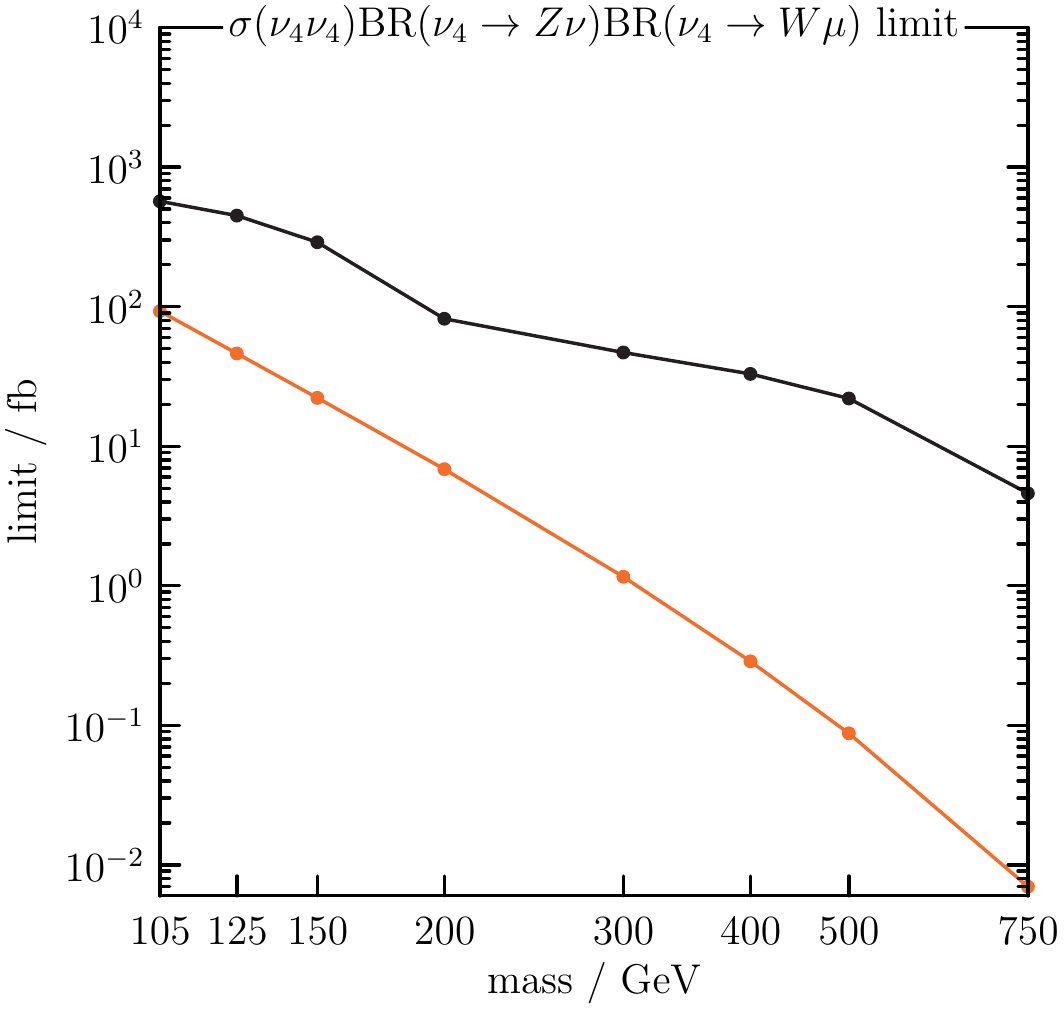}\\
\includegraphics[width=.45\textwidth]{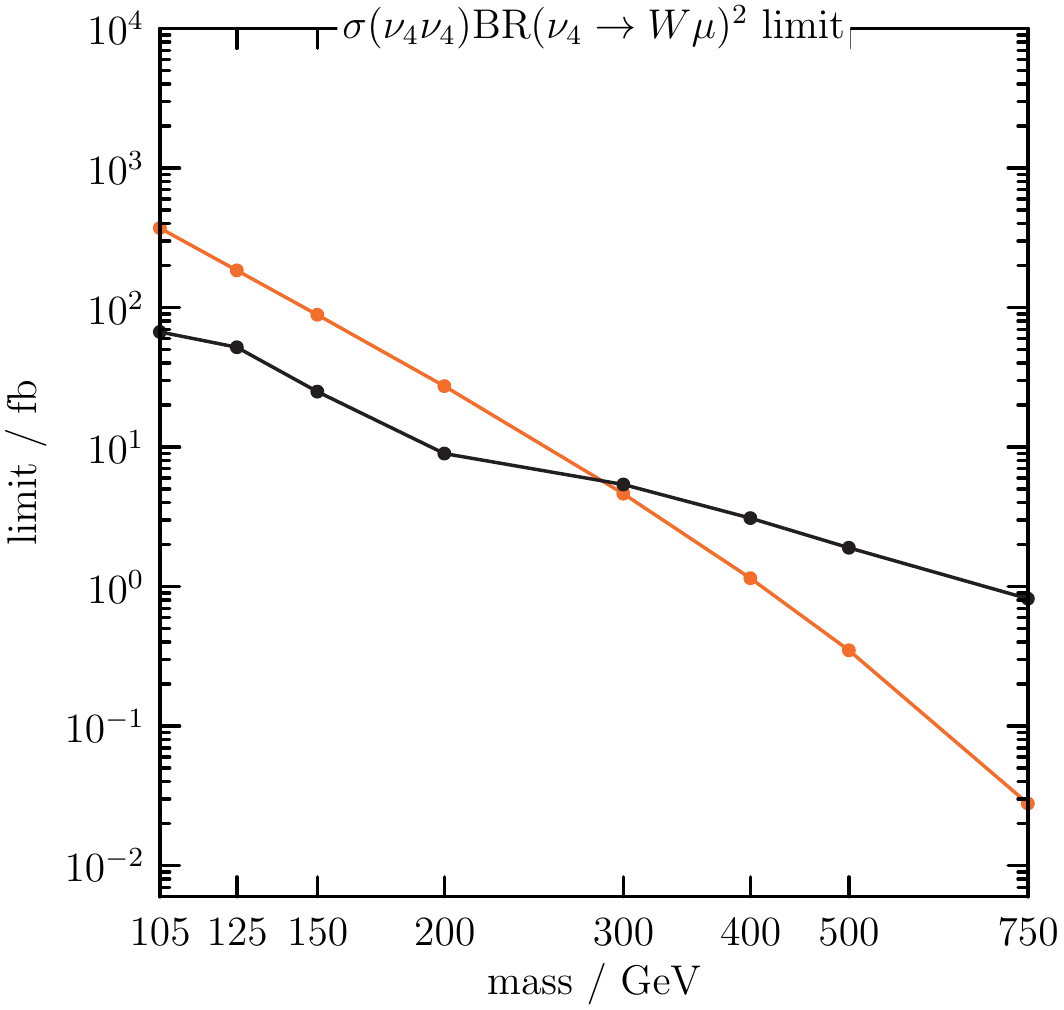}\hfill
\includegraphics[width=.45\textwidth]{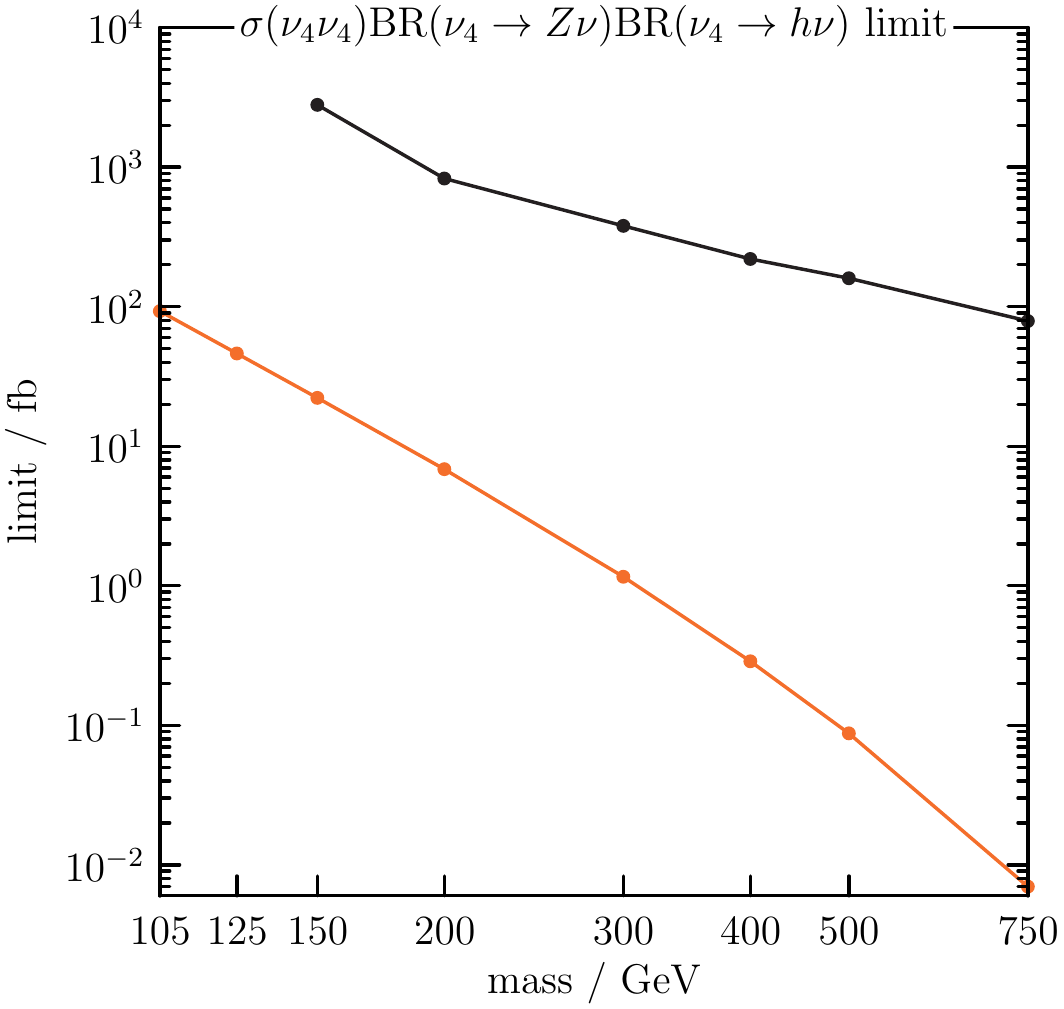}\\
\includegraphics[width=.45\textwidth]{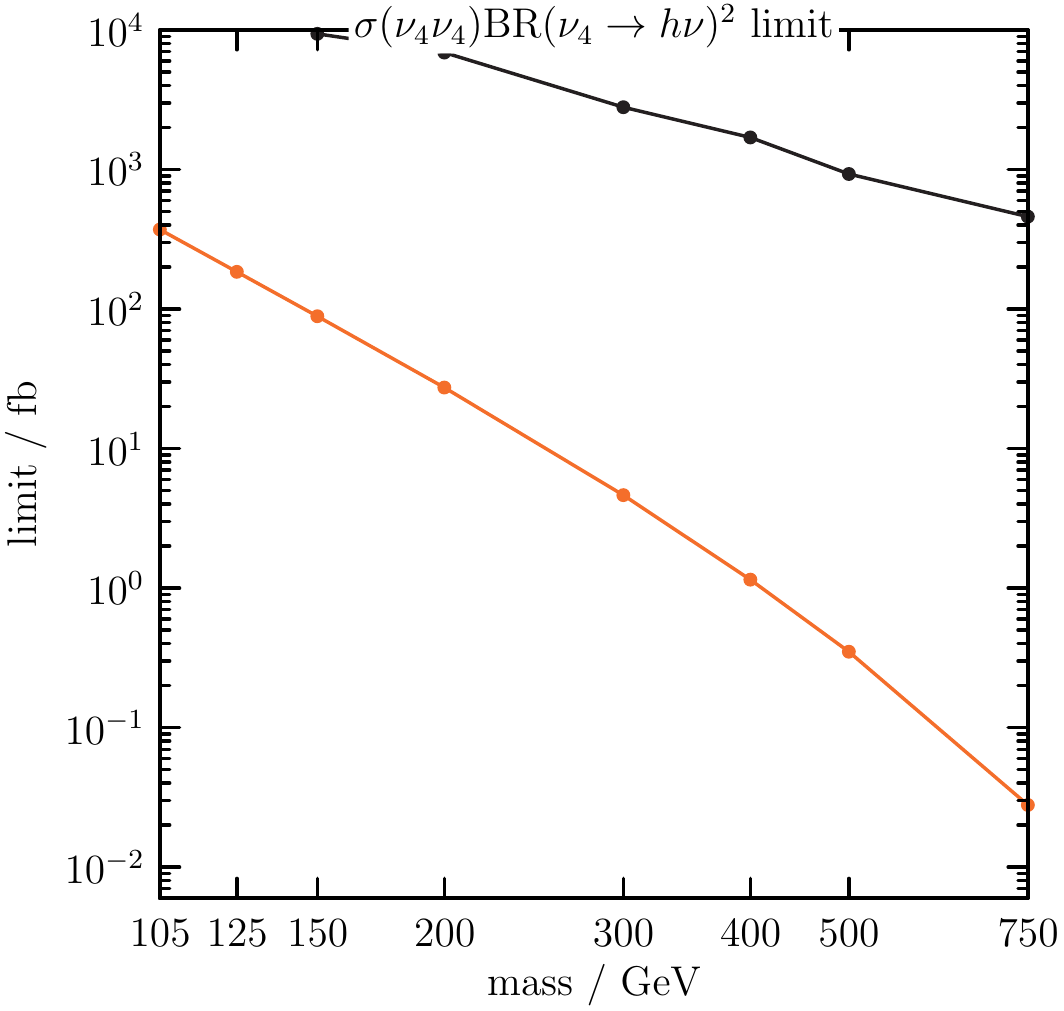}\hfill
\includegraphics[width=.45\textwidth]{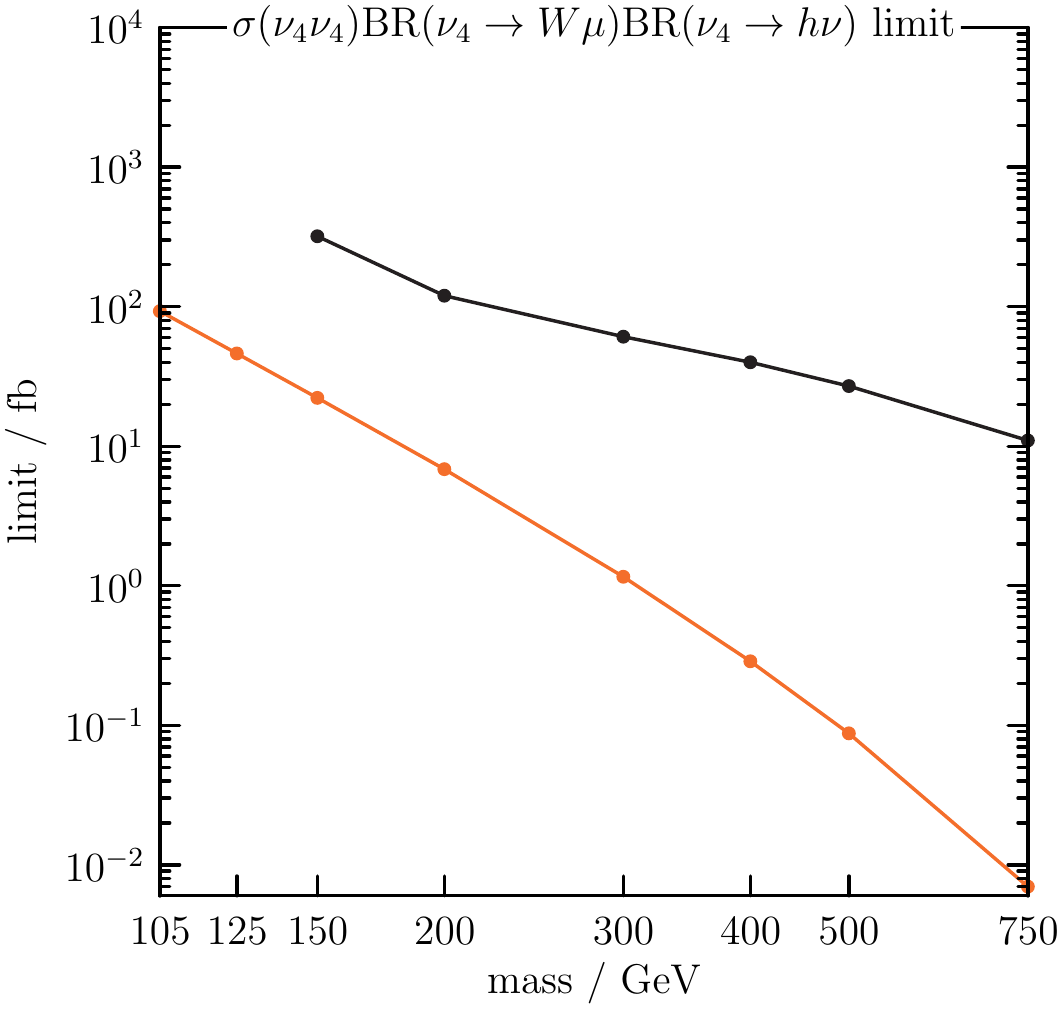}
\caption{The black points show limits on $\sigma(\nu_4 \nu_4)$ times a product of branching ratios for different masses. In the left plots the orange points are the predicted production cross-section $\sigma(\nu_4 \nu_4)$. In the right plots these production cross-sections are multiplied by $\nf{1}{4}$.
\label{fig:NN}}
\end{figure}

The results are presented in figures~\ref{fig:EE}--\ref{fig:NN} and table~\ref{tab:lim}. The black points in the figures are the limits from the $p_T$ cuts in the ATLAS analysis. Grey points show the limits coming from the five-lepton cut in the cases where this provides a better limit (assuming zero data). Green points show the representative limits from the $b$-tag cuts in the cases where they are better. Both the five-lepton cut and the $b$-tag cuts can only compete with the $p_T$ cuts at low mass, the five-lepton cut being effective for processes that can produce that many leptons and the $b$-tag cuts being effective for processes that can produce a Higgs boson plus at least three leptons from elsewhere. However, combining these cuts with the $p_T$ cuts in future analyses could be useful.

The category and cut setting the best limit for each black point is indicated by a two-letter code in the table; this code is explained in table~\ref{tab:code}. The orange points are the predicted Drell-Yan production cross-sections assuming doublet nature; the blue points are the predictions for singlet nature $e_4^+e_4^-$ production. Black points below prediction points indicate that there is a non-trivial constraint on the considered product of branching ratios. Some combinations of branching ratios are poorly constrained, whereas some are constrained up to masses of about 300~GeV; the products of branching ratios $\rr{BR}(e_4 \rightarrow Z\mu)\rr{BR}(\nu_4 \rightarrow W \mu)$ and $\rr{BR}(e_4 \rightarrow h\mu)\rr{BR}(\nu_4 \rightarrow W \mu)$ are constrained up to masses of more than 500~GeV.

Where the limits on a production cross-section times a product of branching ratios is shown compared to one quarter of the predicted cross-section (the right hand plots in figures~\ref{fig:EE} and~\ref{fig:NN}), this is to make it more clear whether the limit on the process actually sets a limit on the branching ratios when the given production cross-section is assumed. If the limit on the product of branching ratios $xy$ of some particle is weaker than $xy < \nf{1}{4}$ then this limit is strictly weaker than the trivial limit $x + y < 1$. If a limit on a given process implies a non-trivial constraint on the relevant branching ratios when doublet production is assumed the entry is coloured orange in the table; if a constraint is also implied for the singlet case the entry is coloured blue.

\begin{table}[tbp]
\centering
\footnotesize
\begin{tabular}{|l|ccccccccc|}
\hline
& \multicolumn{9}{c|}{masses / GeV} \\
\hline
& 105 & 125 & 150 & 200 & 300 & 400 & 500 & 750 & 1000 \\\hline\hline
& \multicolumn{9}{c|}{predicted production cross-sections / fb}
 \\
\hline
$\sigma(e_4^+ e_4^-)$ (singlet) & 426 & 225 & 114 & 37.2 & 6.73 & 1.75 & 0.552 & 0.0481 & 0.00573 \\
$\sigma(e_4^+ e_4^-)$ (doublet) & 1040 & 538 & 269 & 86.6 & 15.5 & 3.98 & 1.24 & 0.106 & 0.0124 \\
$\sigma(e_4^\pm \nu_4)$ (doublet) & 3870 & 1970 & 973 & 310 & 55.5 & 14.4 & 4.53 & 0.378 & 0.0408 \\
$\sigma(\nu_4 \nu_4)$ (doublet) & 372 & 185 & 88.9 & 27.4 & 4.64 & 1.15 & 0.35 & 0.0279 & 0.00306 \\
\hline\hline
& \multicolumn{9}{c|}{95\% C.L. limits / fb and best cuts} \\
\hline
$\sigma(e_4^+ e_4^-) \times$ & \cellcolor{orange!40}530 & \cellcolor{blue!40}190 & \cellcolor{blue!40}66 & \cellcolor{blue!40}21 & \cellcolor{orange!40}12 & 7.5 & 4.8 & 2.2 & 1.9\\
$\rr{BR}(e_4\rightarrow Z\mu)^2$ & \cellcolor{orange!40}{\tt Cb} & \cellcolor{blue!40}{\tt Af} & \cellcolor{blue!40}{\tt Af} & \cellcolor{blue!40}{\tt Af} & \cellcolor{orange!40}{\tt Af} & {\tt Ah} & {\tt Ah} & {\tt Am} & {\tt Am} \\
\hline
$\sigma(e_4^+ e_4^-) \times$ & 520 & 260 & 140 & 65 & 43 & 29 & 23 & 5.1 & 3.7\\
$\rr{BR}(e_4\rightarrow Z\mu)\rr{BR}(e_4\rightarrow W\nu)$ & {\tt Cb} & {\tt Cb} & {\tt Cb} & {\tt Cb} & {\tt Cc} & {\tt Cc} & {\tt Cd} & {\tt Cr} & {\tt Cr} \\
\hline
$\sigma(e_4^+ e_4^-) \times$ & \cellcolor{gray!40} & \cellcolor{gray!40} & 100 & \cellcolor{orange!40}19 & 8.4 & 5.5 & 3.1 & 1.3 & 1.1\\
$\rr{BR}(e_4\rightarrow Z\mu)\rr{BR}(e_4\rightarrow h\mu)$ & \cellcolor{gray!40} & \cellcolor{gray!40} & {\tt Aa} & \cellcolor{orange!40}{\tt Ag} & {\tt Ag} & {\tt Ah} & {\tt Ah} & {\tt Am} & {\tt Am} \\
\hline
$\sigma(e_4^+ e_4^-) \times$ & \cellcolor{gray!40} & \cellcolor{gray!40} & 370 & 130 & 67 & 41 & 28 & 11 & 7.2\\
$\rr{BR}(e_4\rightarrow W\nu)\rr{BR}(e_4\rightarrow h\mu)$ & \cellcolor{gray!40} & \cellcolor{gray!40} & {\tt Ab} & {\tt Ab} & {\tt Ab} & {\tt Ac} & {\tt Ac} & {\tt Am} & {\tt Am} \\
\hline
$\sigma(e_4^+ e_4^-) \times$ & \cellcolor{gray!40} & \cellcolor{gray!40} & \cellcolor{orange!40}220 & \cellcolor{orange!40}64 & 17 & 14 & 7.2 & 2.5 & 2.1\\
$\rr{BR}(e_4\rightarrow h\mu)^2$ & \cellcolor{gray!40} & \cellcolor{gray!40} & \cellcolor{orange!40}{\tt Aa} & \cellcolor{orange!40}{\tt Ag} & {\tt Ag} & {\tt Ag} & {\tt Ah} & {\tt Am} & {\tt Am} \\
\hline
$\sigma(e_4^\pm \nu_4) \times$ & \cellcolor{orange!40}820 & \cellcolor{orange!40}510 & \cellcolor{orange!40}230 & \cellcolor{orange!40}79 & \cellcolor{orange!40}44 & 29 & 23 & 4.8 & 3.4\\
$\rr{BR}(e_4\rightarrow Z\mu)\rr{BR}(\nu_4\rightarrow Z\nu)$ & \cellcolor{orange!40}{\tt Cb} & \cellcolor{orange!40}{\tt Cb} & \cellcolor{orange!40}{\tt Cb} & \cellcolor{orange!40}{\tt Cb} & \cellcolor{orange!40}{\tt Cb} & {\tt Cc} & {\tt Cd} & {\tt Cr} & {\tt Cr} \\
\hline
$\sigma(e_4^\pm \nu_4) \times$ & \cellcolor{orange!40}190 & \cellcolor{orange!40}83 & \cellcolor{orange!40}45 & \cellcolor{orange!40}13 & \cellcolor{orange!40}7.3 & \cellcolor{orange!40}4.7 & \cellcolor{orange!40}2.8 & 1.2 & 1\\
$\rr{BR}(e_4\rightarrow Z\mu)\rr{BR}(\nu_4\rightarrow W\mu)$ & \cellcolor{orange!40}{\tt Aa} & \cellcolor{orange!40}{\tt Aa} & \cellcolor{orange!40}{\tt Ag} & \cellcolor{orange!40}{\tt Ag} & \cellcolor{orange!40}{\tt Af} & \cellcolor{orange!40}{\tt Ah} & \cellcolor{orange!40}{\tt Ah} & {\tt Am} & {\tt Am} \\
\hline
$\sigma(e_4^\pm \nu_4) \times$ & \cellcolor{orange!40}2700 & \cellcolor{orange!40}1800 & 1100 & 520 & 330 & 150 & 110 & 45 & 42\\
$\rr{BR}(e_4\rightarrow W\nu)\rr{BR}(\nu_4\rightarrow Z\nu)$ & \cellcolor{orange!40}{\tt Cb} & \cellcolor{orange!40}{\tt Cb} & {\tt Cb} & {\tt Cb} & {\tt Cb} & {\tt Cc} & {\tt Cd} & {\tt Cd} & {\tt Cd} \\
\hline
$\sigma(e_4^\pm \nu_4) \times$ & \cellcolor{orange!40}420 & \cellcolor{orange!40}400 & \cellcolor{orange!40}260 & \cellcolor{orange!40}110 & 57 & 32 & 21 & 11 & 7.1\\
$\rr{BR}(e_4\rightarrow W\nu)\rr{BR}(\nu_4\rightarrow W\mu)$ & \cellcolor{orange!40}{\tt Aa} & \cellcolor{orange!40}{\tt Aa} & \cellcolor{orange!40}{\tt Ab} & \cellcolor{orange!40}{\tt Ag} & {\tt Ab} & {\tt Ac} & {\tt Ac} & {\tt Am} & {\tt Am} \\
\hline
$\sigma(e_4^\pm \nu_4) \times$ & \cellcolor{gray!40} & \cellcolor{gray!40} & 1100 & \cellcolor{orange!40}280 & 110 & 64 & 51 & 9.8 & 7.7\\
$\rr{BR}(e_4\rightarrow Z\mu)\rr{BR}(\nu_4\rightarrow h\nu)$ & \cellcolor{gray!40} & \cellcolor{gray!40} & {\tt Aa} & \cellcolor{orange!40}{\tt Cb} & {\tt Cb} & {\tt Cc} & {\tt Cr} & {\tt Cr} & {\tt Cr} \\
\hline
$\sigma(e_4^\pm \nu_4) \times$ & \cellcolor{gray!40} & \cellcolor{gray!40} & 1400 & \cellcolor{orange!40}250 & 110 & 75 & 53 & 9.3 & 7.1\\
$\rr{BR}(e_4\rightarrow h\mu)\rr{BR}(\nu_4\rightarrow Z\nu)$ & \cellcolor{gray!40} & \cellcolor{gray!40} & {\tt Aa} & \cellcolor{orange!40}{\tt Cb} & {\tt Cb} & {\tt Cq} & {\tt Cr} & {\tt Cr} & {\tt Cr} \\
\hline
$\sigma(e_4^\pm \nu_4) \times$ & \cellcolor{gray!40} & \cellcolor{gray!40} & 6400 & 5000 & 1800 & 1200 & 680 & 360 & 270\\
$\rr{BR}(e_4\rightarrow W\nu)\rr{BR}(\nu_4\rightarrow h\nu)$ & \cellcolor{gray!40} & \cellcolor{gray!40} & {\tt Ab} & {\tt Ap} & {\tt Ab} & {\tt Bc} & {\tt Ac} & {\tt Ac} & {\tt Bc} \\
\hline
$\sigma(e_4^\pm \nu_4) \times$ & \cellcolor{gray!40} & \cellcolor{gray!40} & \cellcolor{orange!40}110 & \cellcolor{orange!40}20 & \cellcolor{orange!40}9.2 & \cellcolor{orange!40}6.3 & \cellcolor{orange!40}3.5 & 1.5 & 1.2\\
$\rr{BR}(e_4\rightarrow h\mu)\rr{BR}(\nu_4\rightarrow W\mu)$ & \cellcolor{gray!40} & \cellcolor{gray!40} & \cellcolor{orange!40}{\tt Aa} & \cellcolor{orange!40}{\tt Ag} & \cellcolor{orange!40}{\tt Ag} & \cellcolor{orange!40}{\tt Ah} & \cellcolor{orange!40}{\tt Ah} & {\tt Am} & {\tt Am} \\
\hline
$\sigma(e_4^\pm \nu_4) \times$ & \cellcolor{gray!40} & \cellcolor{gray!40} & \cellcolor{orange!40}910 & 420 & 140 & 93 & 52 & 19 & 13\\
$\rr{BR}(e_4\rightarrow h\mu)\rr{BR}(\nu_4\rightarrow h\nu)$ & \cellcolor{gray!40} & \cellcolor{gray!40} & \cellcolor{orange!40}{\tt Aa} & {\tt Ap} & {\tt Ap} & {\tt Aq} & {\tt An} & {\tt Am} & {\tt Am} \\
\hline
$\sigma(\nu_4 \nu_4) \times$ & 5100 & 5700 & 4000 & 850 & 450 & 200 & 150 & 87 & 73\\
$\rr{BR}(\nu_4\rightarrow Z\nu)^2$ & {\tt Cc} & {\tt Cf} & {\tt Cb} & {\tt Cb} & {\tt Cc} & {\tt Cc} & {\tt Cd} & {\tt Cd} & {\tt Cd} \\
\hline
$\sigma(\nu_4 \nu_4) \times$ & 570 & 450 & 290 & 82 & 47 & 33 & 22 & 4.6 & 3.5\\
$\rr{BR}(\nu_4\rightarrow Z\nu)\rr{BR}(\nu_4\rightarrow W\mu)$ & {\tt Ag} & {\tt Ag} & {\tt Ag} & {\tt Cb} & {\tt Cb} & {\tt Cc} & {\tt Cr} & {\tt Cr} & {\tt Cr} \\
\hline
$\sigma(\nu_4 \nu_4) \times$ & \cellcolor{orange!40}67 & \cellcolor{orange!40}52 & \cellcolor{orange!40}25 & \cellcolor{orange!40}9 & 5.4 & 3.1 & 1.9 & 0.82 & 0.72\\
$\rr{BR}(\nu_4\rightarrow W\mu)^2$ & \cellcolor{orange!40}{\tt Aa} & \cellcolor{orange!40}{\tt Aa} & \cellcolor{orange!40}{\tt Ag} & \cellcolor{orange!40}{\tt Ag} & {\tt Af} & {\tt Ah} & {\tt Am} & {\tt Am} & {\tt Am} \\
\hline
$\sigma(\nu_4 \nu_4) \times$ & \cellcolor{gray!40} & \cellcolor{gray!40} & 2800 & 830 & 380 & 220 & 160 & 79 & 72\\
$\rr{BR}(\nu_4\rightarrow Z\nu)\rr{BR}(\nu_4\rightarrow h\nu)$ & \cellcolor{gray!40} & \cellcolor{gray!40} & {\tt Cb} & {\tt Cb} & {\tt Cb} & {\tt Cc} & {\tt Cc} & {\tt Cd} & {\tt Cd} \\
\hline
$\sigma(\nu_4 \nu_4) \times$ & \cellcolor{gray!40} & \cellcolor{gray!40} & 320 & 120 & 61 & 40 & 27 & 11 & 6.9\\
$\rr{BR}(\nu_4\rightarrow W\mu)\rr{BR}(\nu_4\rightarrow h\nu)$ & \cellcolor{gray!40} & \cellcolor{gray!40} & {\tt Ag} & {\tt Ag} & {\tt Ag} & {\tt Ac} & {\tt Ac} & {\tt Am} & {\tt Am} \\
\hline
$\sigma(\nu_4 \nu_4) \times$ & \cellcolor{gray!40} & \cellcolor{gray!40} & 9400 & 6900 & 2800 & 1700 & 930 & 460 & 380\\
$\rr{BR}(\nu_4\rightarrow h\nu)^2$ & \cellcolor{gray!40} & \cellcolor{gray!40} & {\tt Aa} & {\tt Ap} & {\tt Ab} & {\tt Bc} & {\tt Bc} & {\tt Bc} & {\tt Bc} \\
\hline
\end{tabular}

\caption{The predicted production cross-sections and limits on the various cross-sections times products of branching ratios for different heavy lepton masses. Orange limits imply a non-trivial limit on the product of branching ratios assuming doublet production; blue limits additionally imply a non-trivial limit on the product of branching ratios assuming singlet production. The category and cut setting the best limit is indicated by a two-letter code; this code is explained in table~\ref{tab:code}.
\label{tab:lim}}
\end{table}

\begin{table}[tbp]
\centering
\footnotesize
\begin{tabular}{|c|l|}
\hline
\cellcolor{blue!40}{\tt A} & \cellcolor{blue!40}$\ge$3$e/\mu$ off-$Z$ \\
{\tt B} & 2$e/\mu + \tau_h$ off-$Z$ \\
\cellcolor{orange!40}{\tt C} & \cellcolor{orange!40}$\ge$3$e/\mu$ on-$Z$ \\
{\tt D} & 2$e/\mu + \tau_h$ on-$Z$ \\
\hline
\cellcolor{orange!40}{\tt a} & \cellcolor{orange!40}$H_T^j < 150$ GeV \\
\cellcolor{orange!40}{\tt b} & \cellcolor{orange!40}$H_T^j < 150$ GeV, $\ff{E}_T > 100$ GeV \\
{\tt c} & $H_T^j < 150$ GeV, $\ff{E}_T > 200$ GeV \\
{\tt d} & $H_T^j < 150$ GeV, $\ff{E}_T > 300$ GeV \\
\cellcolor{blue!40}{\tt f} & \cellcolor{blue!40}min $p_T^l > 50$ GeV \\
\cellcolor{orange!40}{\tt g} & \cellcolor{orange!40}$H_T^l > 200$ GeV \\
\cellcolor{orange!40}{\tt h} & \cellcolor{orange!40}$H_T^l > 500$ GeV \\
{\tt m} & $m_{\rr{eff}} > 1000$ GeV \\
{\tt n} & $H_T^j > 150$ GeV, $\ff{E}_T > 200$ GeV \\
{\tt p} & $\ff{E}_T > 100$ GeV \\
{\tt q} & $\ff{E}_T > 100$ GeV, $m_{\rr{eff}} > 600$ GeV \\
{\tt r} & $\ff{E}_T > 100$ GeV, $m_{\rr{eff}} > 1200$ GeV \\
\hline
\end{tabular}
\caption{The code letters for the categories and cuts. The colours are from table~\ref{tab:lim} (where this code is used) and indicate that the coloured category or cut is useful for setting a limit on branching ratios in this analysis. Uncoloured categories and cuts (except {\tt D}) {\em do} set best limits on cross-sections times products of branching ratios in this analysis, but apply at too high masses (or to too difficult-to-observe processes) to be useful with current data.
\label{tab:code}}
\end{table}

\subsection{Combined results under different assumptions}

Let us now compare the limits with predicted production cross-sections in order to derive limits on branching ratios in the different scenarios (summarised in table~\ref{tab:ver}). In the singlet case the only limits on the branching ratios are those coming from the limits on $\sigma(e_4^+ e_4^-)\rr{BR}(e_4\rightarrow Z\mu)^2$. They say that at 125, 150, and 200~GeV BR($e_4\rightarrow Z\mu$) must be less than 92\%, 76\%, and 75\% respectively. There are, however, no limits at 105~GeV (and from 300~GeV upwards).\footnote{Note that bounds obtained from CMS data for this scenario (see Ref.~\cite{Falkowski:2013jya}) have been derived under the assumption of ${\rm BR}(e_4\rightarrow Z\mu) \sim 40\%$ and appear to be stronger than ours at very low vectorlike lepton masses.}

\begin{figure}[tbp]
\centering
\includegraphics[width=.7\textwidth]{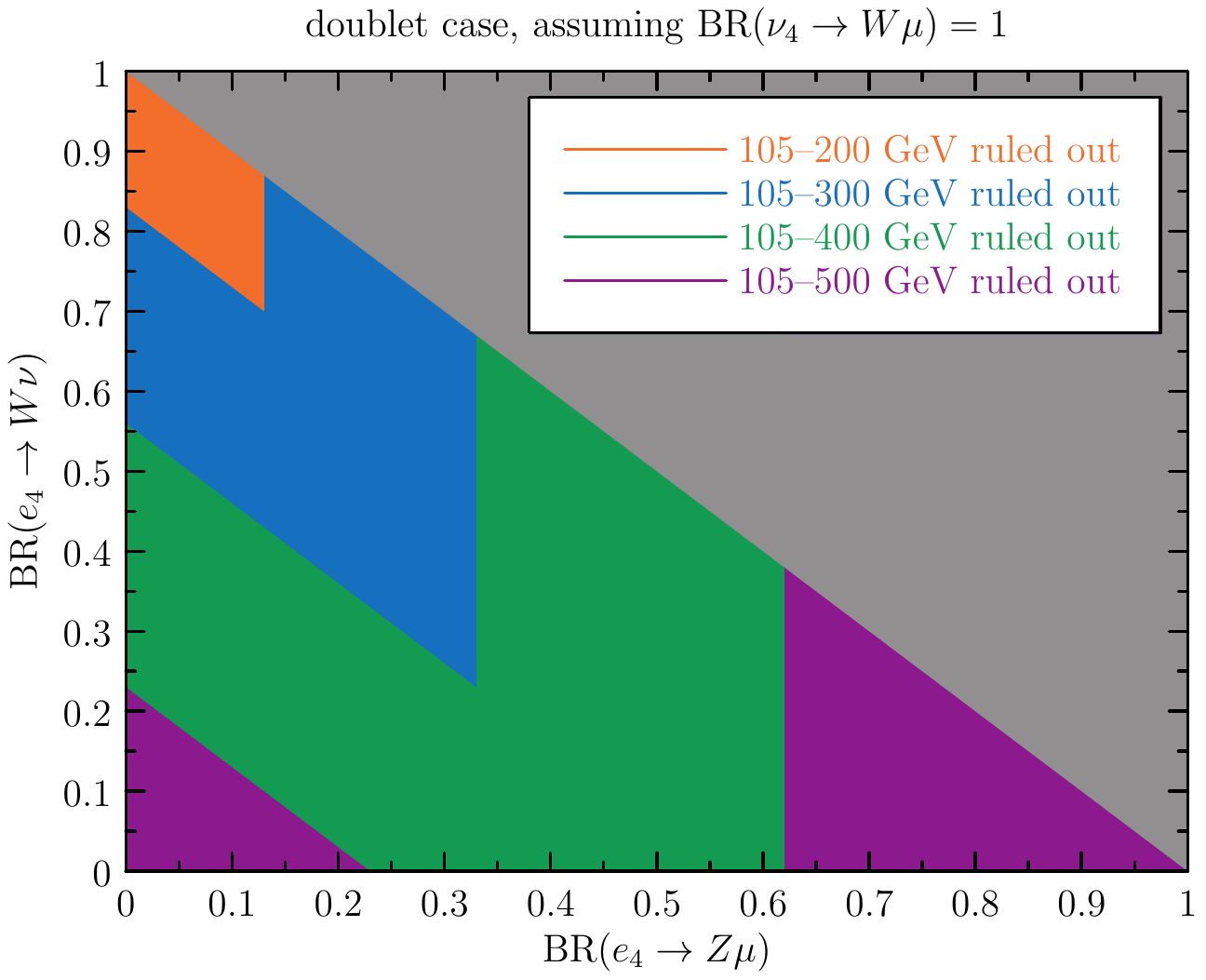}
\caption{The doublet case assuming $\rr{BR}(\nu_4\rightarrow W\mu) = 1$. Heavy lepton masses of 200~GeV and below are {\em completely} ruled out. From 300--500~GeV the limits on $\rr{BR}(e_4\rightarrow Z\mu)\rr{BR}(\nu_4\rightarrow W\mu) = \rr{BR}(e_4\rightarrow Z\mu)$ push in from the right and the limits on $\rr{BR}(e_4\rightarrow h\mu)\rr{BR}(\nu_4\rightarrow W\mu) = \rr{BR}(e_4\rightarrow h\mu)$ push in from the bottom-left.
\label{fig:ra}}
\end{figure}

In the doublet cases where it is assumed that $\rr{BR}(\nu_4\rightarrow W\mu) = 1$ strong bounds can be set. The limits on $\sigma(\nu_4 \nu_4)\rr{BR}(\nu_4\rightarrow W\mu)^2$ instantly rule out all masses below about 300~GeV. Then there are strong constraints coming from the limits on $\sigma(e_4^\pm \nu_4)\rr{BR}(e_4\rightarrow Z\mu)\rr{BR}(\nu_4\rightarrow W\mu)$ and $\sigma(e_4^\pm \nu_4)\rr{BR}(e_4\rightarrow h\mu)\rr{BR}(\nu_4\rightarrow W\mu)$, constraining branching ratios beyond 500~GeV. The situation is summarised in figure~\ref{fig:ra}, where the limits are shown in the $[{\rm BR} (e_4\to Z \mu), {\rm BR} (e_4 \to W \nu)]$ plane; in this case these are the only two independent parameters.

\begin{figure}[tbp]
\centering
\includegraphics[width=.45\textwidth]{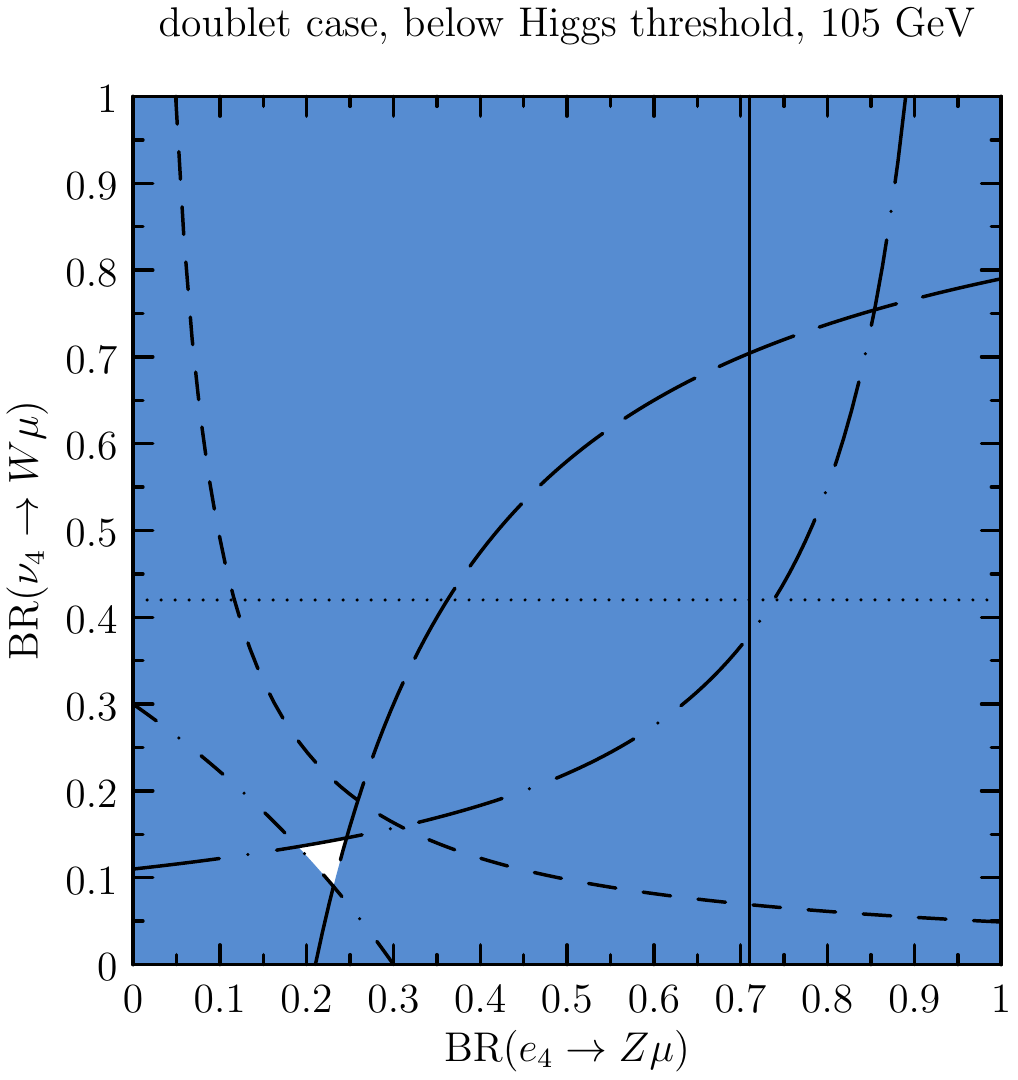}\hfill
\includegraphics[width=.45\textwidth]{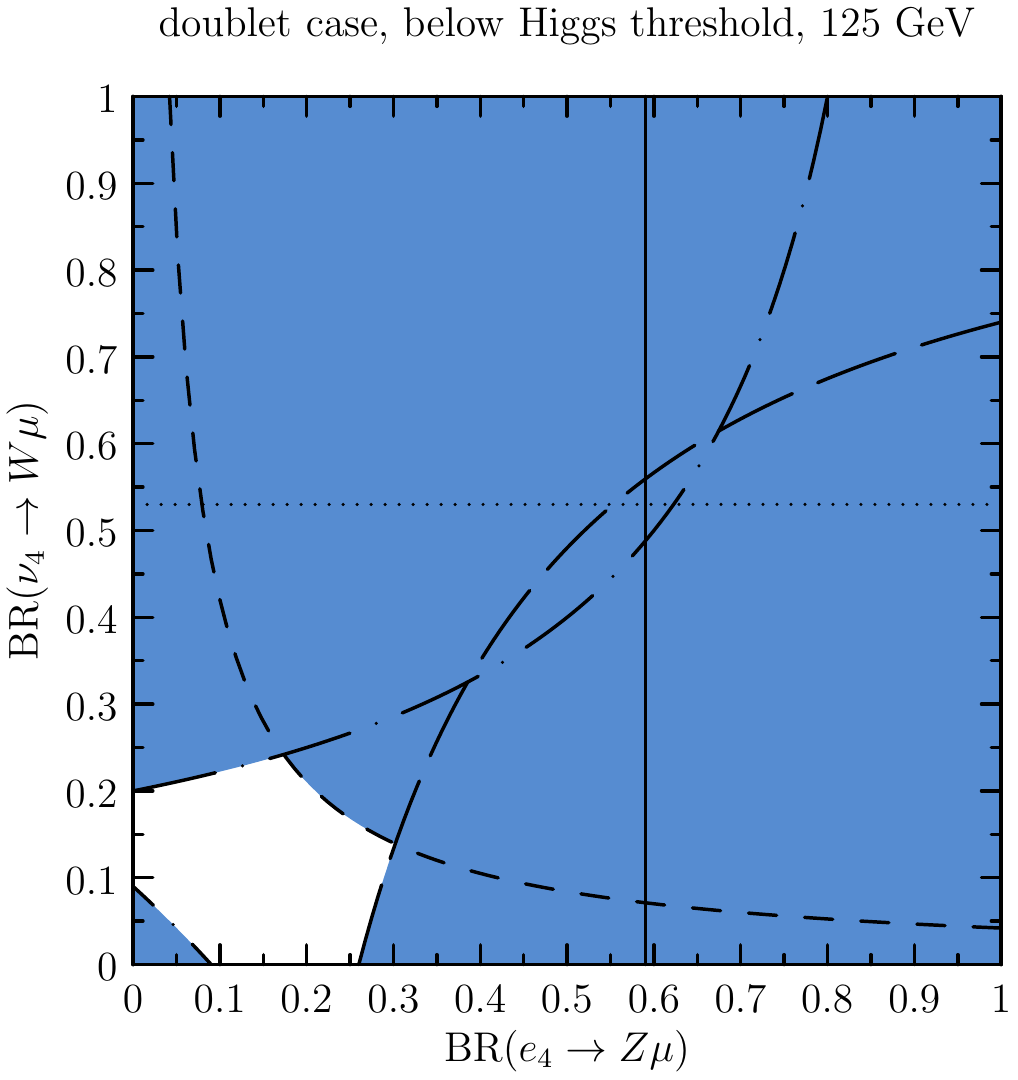}
\caption{The general doublet case below the Higgs threshold. Blue points are ruled out; white points are allowed.
The limits on $\rr{BR}(e_4\rightarrow Z\mu)$ (solid) and $\rr{BR}(\nu_4\rightarrow W\mu)$ (dotted) push in from the right and top respectively. The limits on $\rr{BR}(e_4\rightarrow Z\mu)\rr{BR}(\nu_4\rightarrow W\mu)$ (short-dashed) push in from the top-right whereas the limits on $\rr{BR}(e_4\rightarrow W\nu)\rr{BR}(\nu_4\rightarrow Z\nu)$ (short-dash-dotted) push in from the bottom-left. The limits on $\rr{BR}(e_4\rightarrow Z\mu)\rr{BR}(\nu_4\rightarrow Z\nu)$ (long-dashed) and $\rr{BR}(e_4\rightarrow W\nu)\rr{BR}(\nu_4\rightarrow W\mu)$ (long-dash-dotted) push in from the bottom-right and top-left respectively.
\label{fig:db}}
\end{figure}

When the condition $\rr{BR}(\nu_4\rightarrow W\mu) = 1$ is relaxed the situation becomes more complicated, but below the Higgs threshold we can still represent the limits in a two-dimensional parameter space. In this case the situation is quite constrained due to many limits constraining in different directions. The ruled out parameter space at 105 and 125~GeV is shown in figure~\ref{fig:db}. The allowed space is in white. If the five-lepton cut results are valid (there is indeed zero data) then the hole in the 105~GeV case closes completely and even in the 125~GeV case only the part of the allowed region to the left of about 0.1 in  ${\rm BR} (e_4\to Z \mu)$ remains.

\begin{figure}[tbp]
\centering
\includegraphics[width=.45\linewidth]{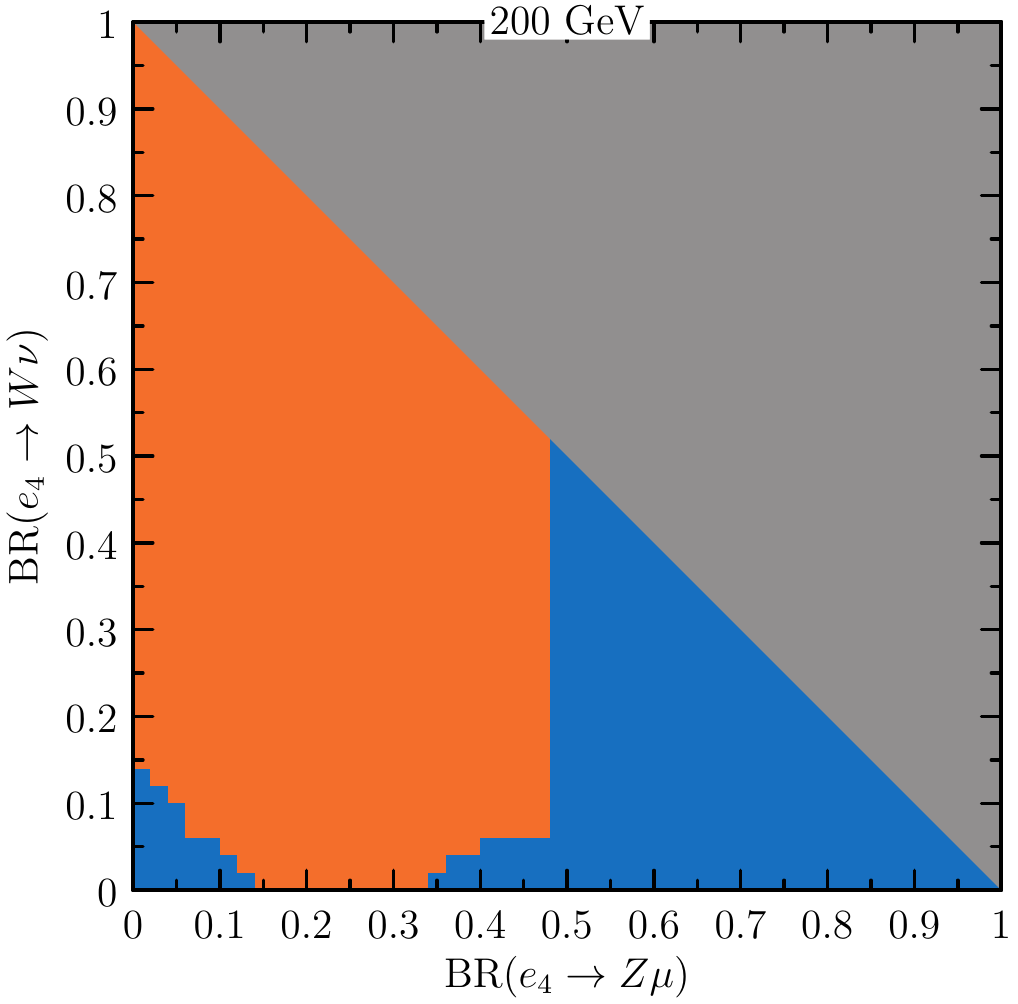}\hfill
\includegraphics[width=.45\linewidth]{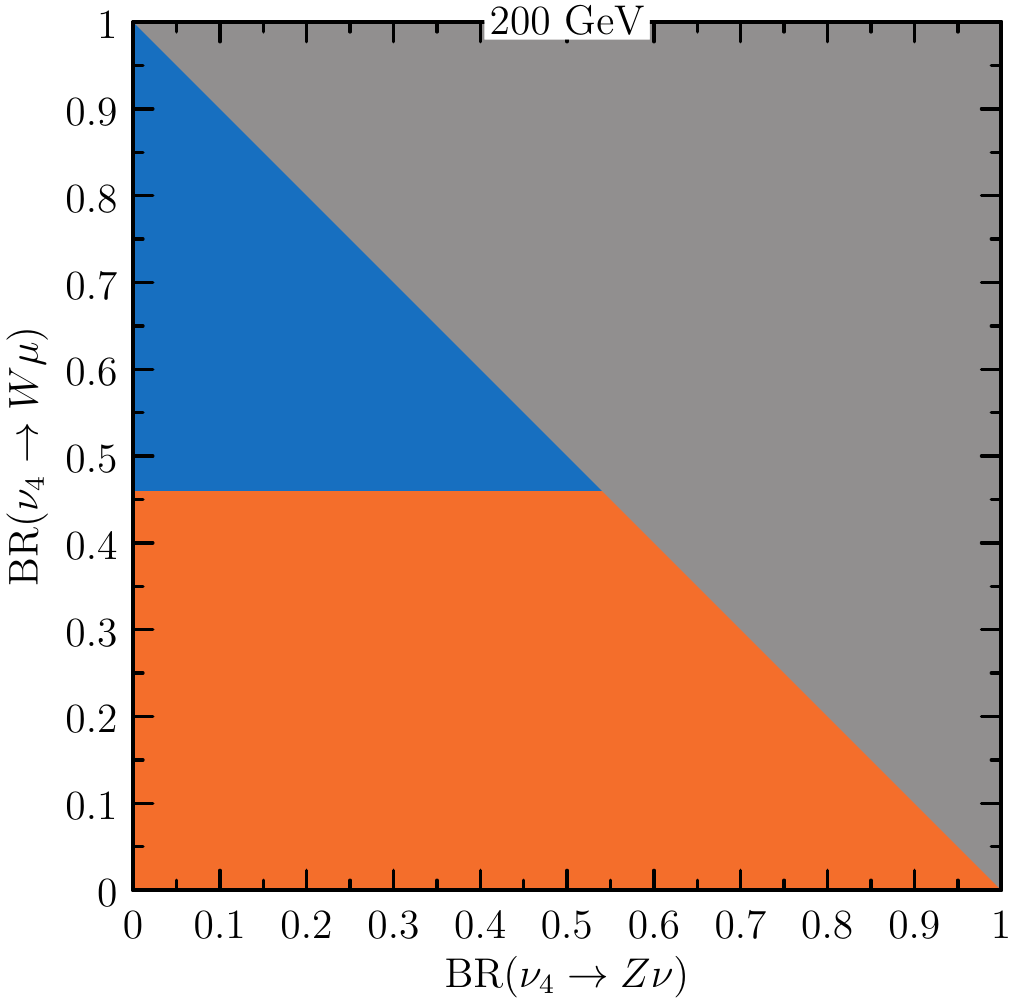}\\
\includegraphics[width=.45\linewidth]{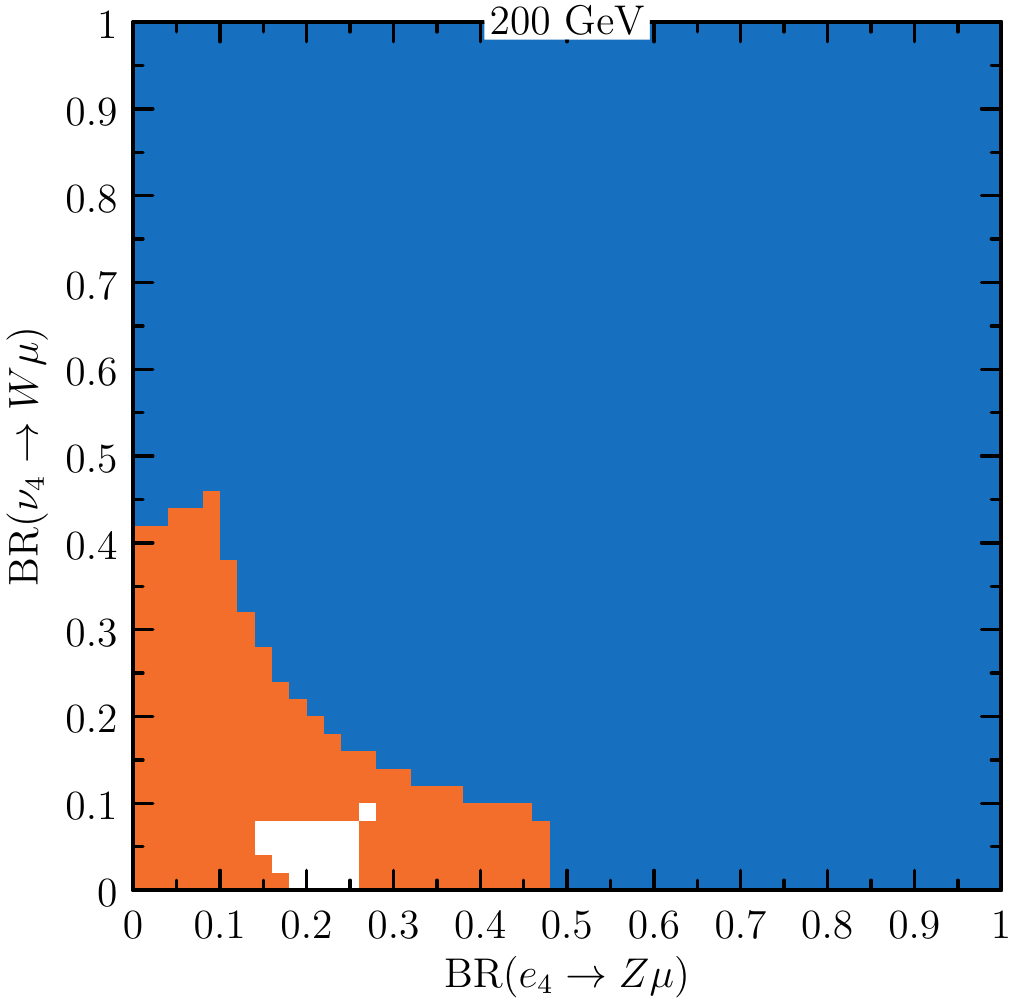}\hfill
\includegraphics[width=.45\linewidth]{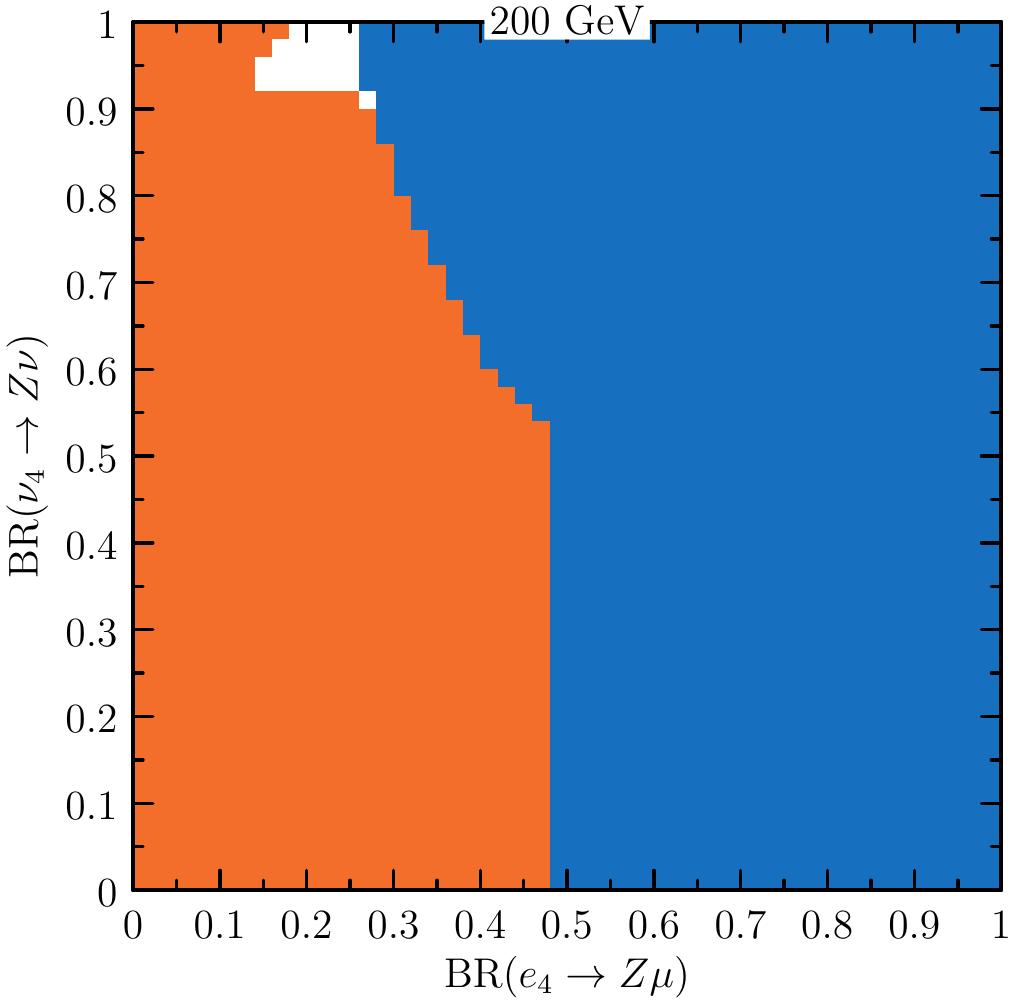}\\
\includegraphics[width=.45\linewidth]{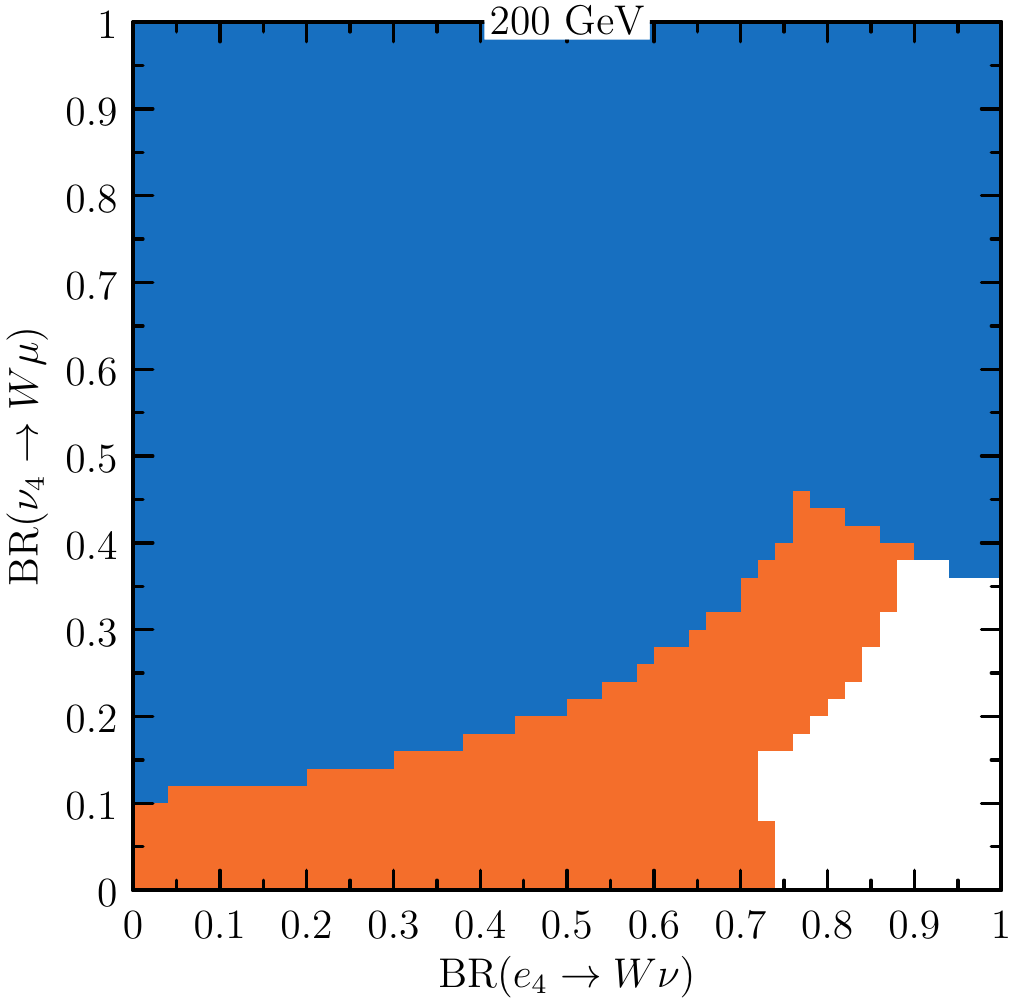}\hfill
\includegraphics[width=.45\linewidth]{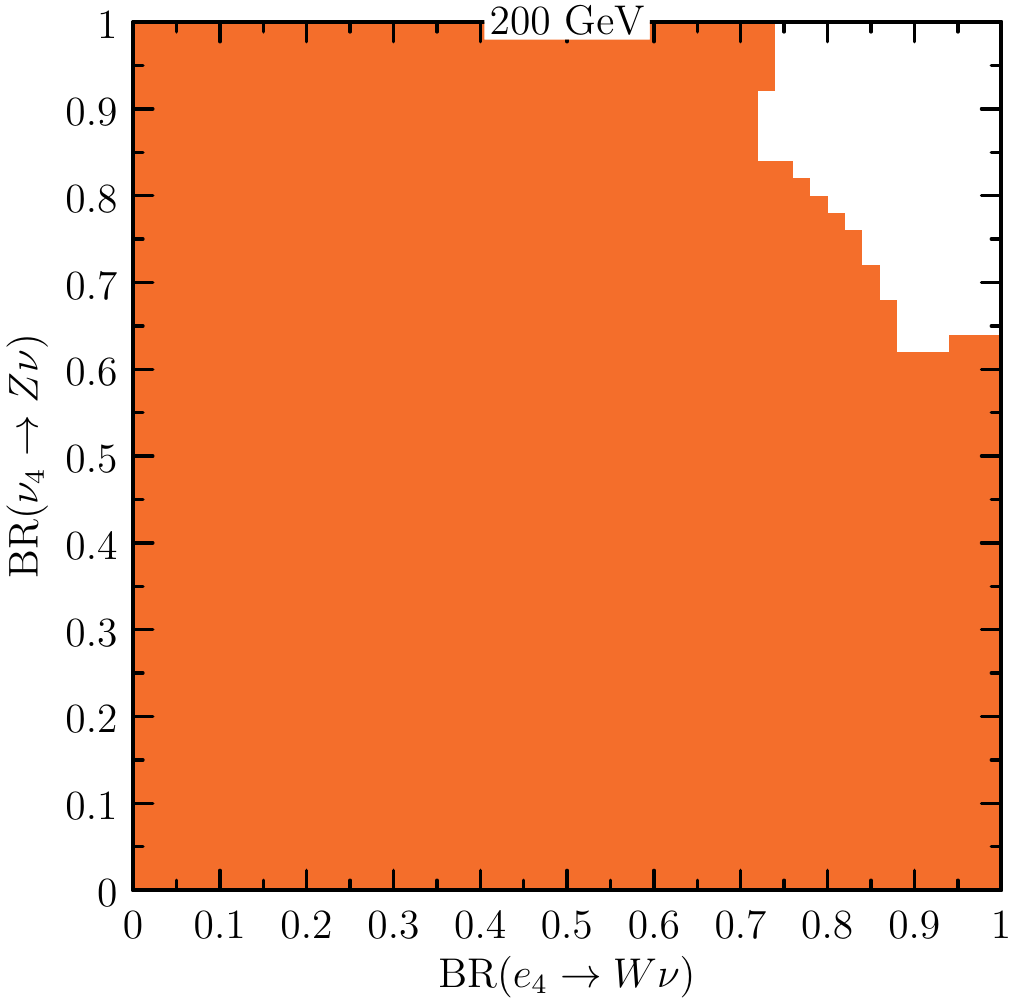}
\caption{The general doublet case for 200~GeV---above the Higgs threshold. Blue points are ruled out for all values of the other two independent branching ratios; white points are allowed for all values of the other two independent branching ratios; other points are shown in orange. These are the six combinations of the four independent branching ratios.
\label{fig:da200}}
\end{figure}

\begin{figure}[tbp]
\centering
\includegraphics[width=.45\linewidth]{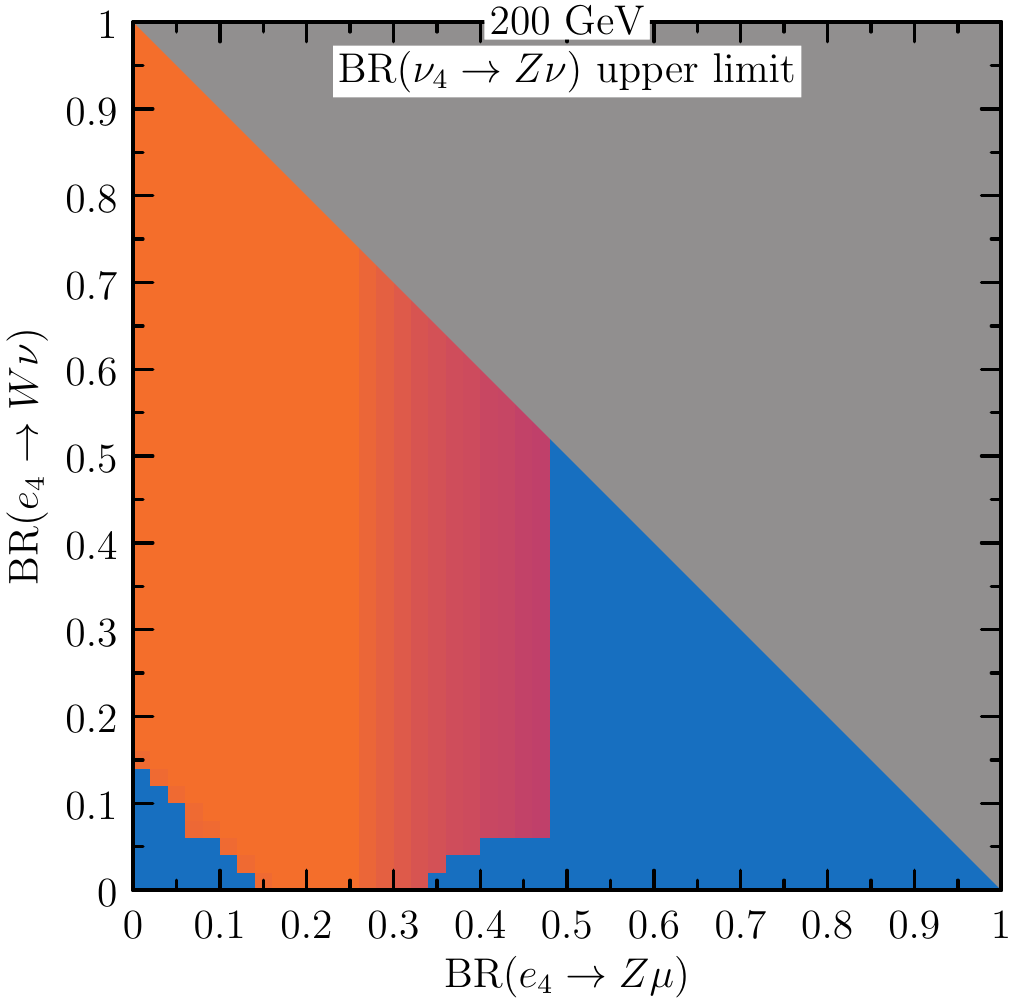}\hfill
\includegraphics[width=.45\linewidth]{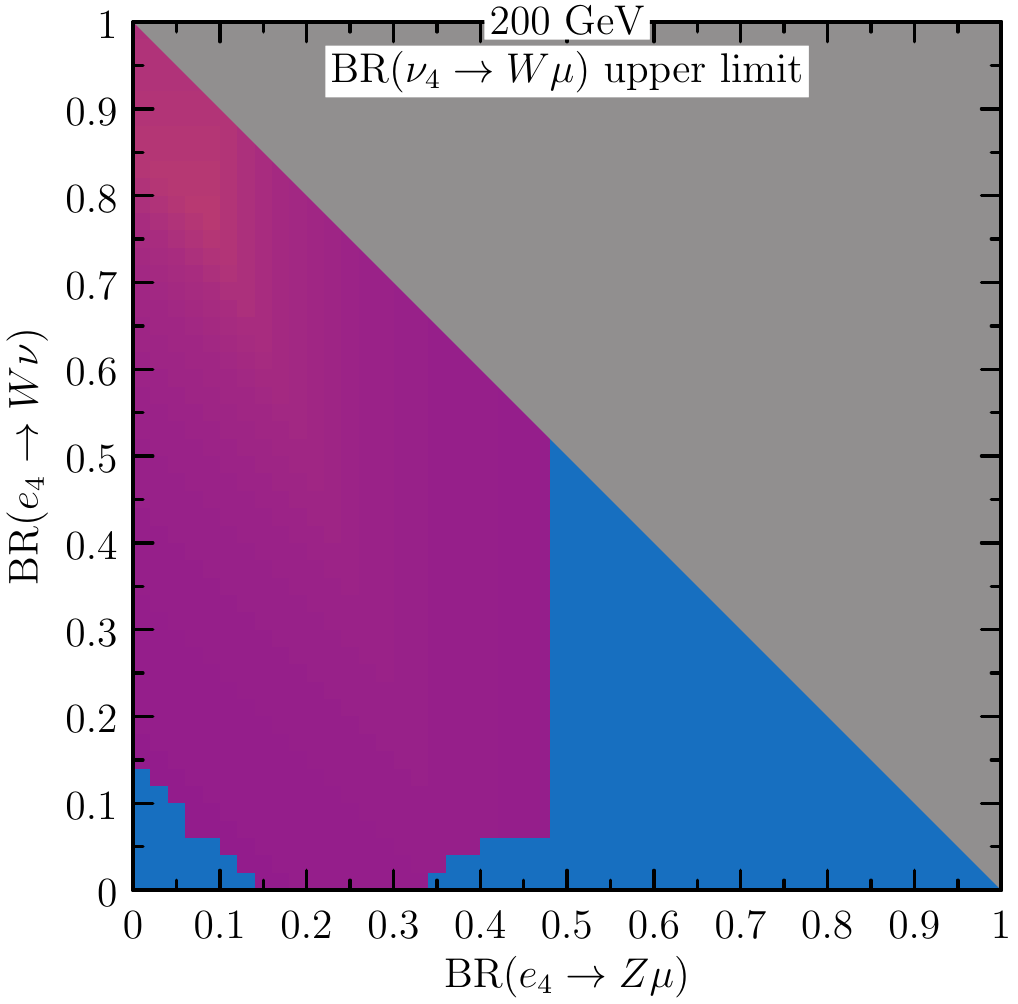}\\
\includegraphics[width=.45\linewidth]{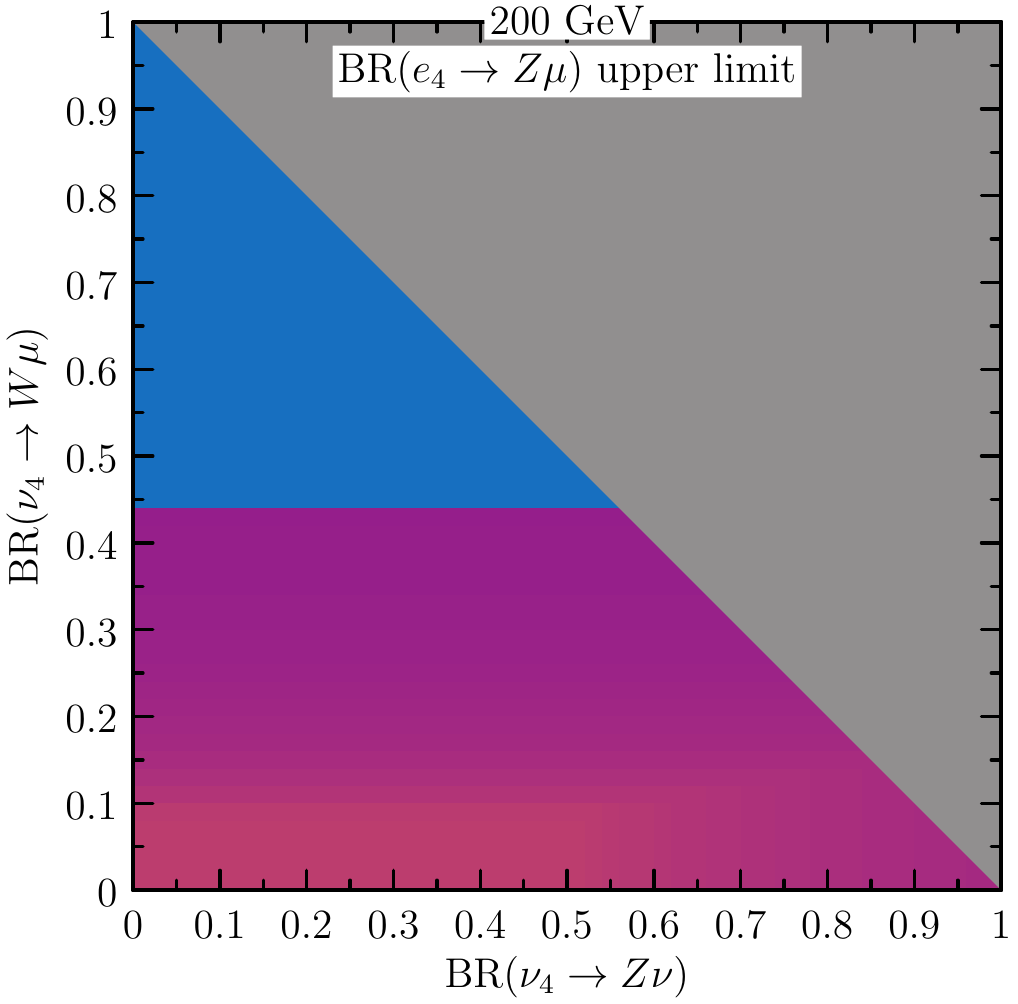}\hfill
\includegraphics[width=.45\linewidth]{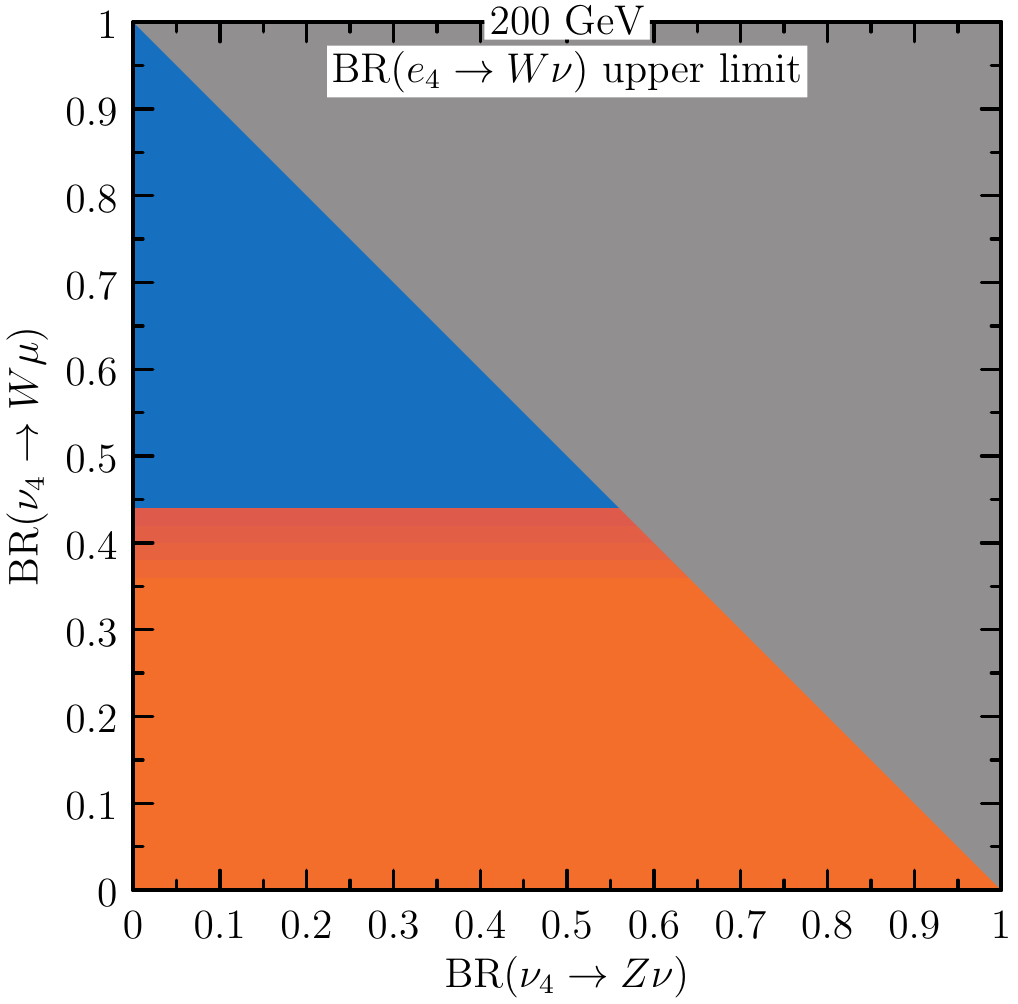}\\
\raisebox{-0.5\height}{\includegraphics[width=.45\linewidth]{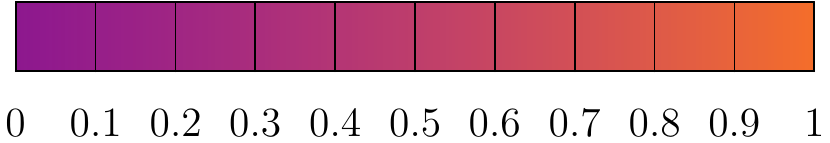}}\hfill
\raisebox{-0.5\height}{\includegraphics[width=.45\linewidth]{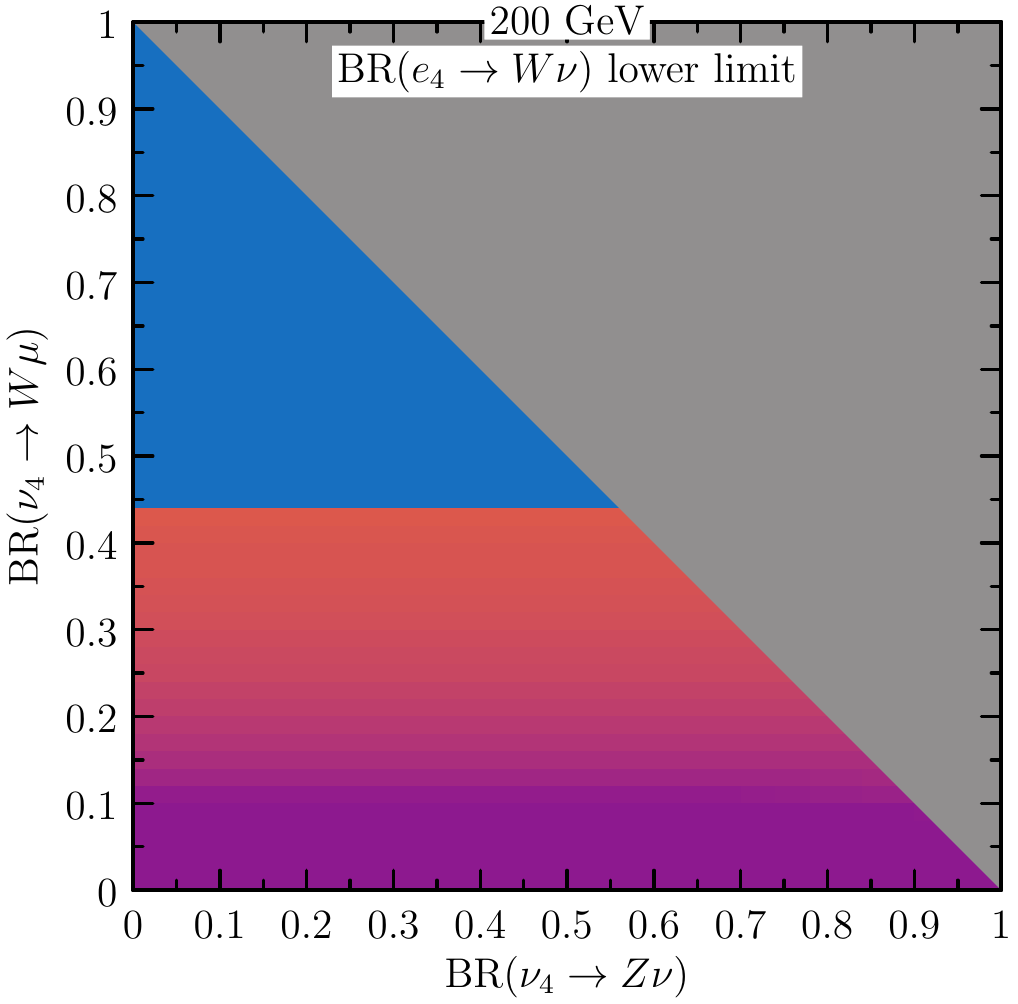}}
\caption{The general doublet case for 200~GeV. A more detailed look at the top two projections of figure~\ref{fig:da200}. Blue points are ruled out for all values of the other two independent branching ratios; white points are allowed for all values of the other two independent branching ratios; other points are shown in orange/purple. The orange/purple gradient shows the upper or lower limit on the a third independent branching ratio.
\label{fig:cont200}}
\end{figure}

\begin{figure}[tbp]
\centering
\includegraphics[width=.45\linewidth]{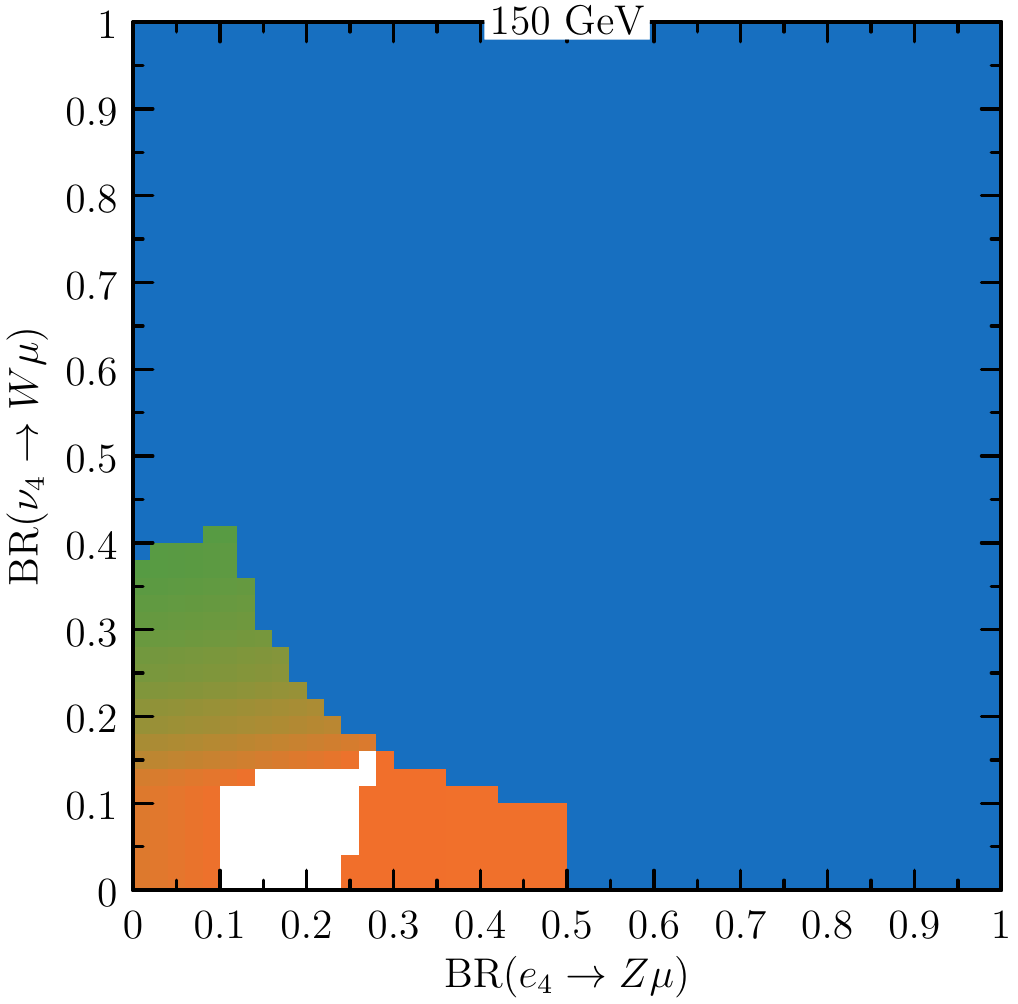}\hfill
\includegraphics[width=.45\linewidth]{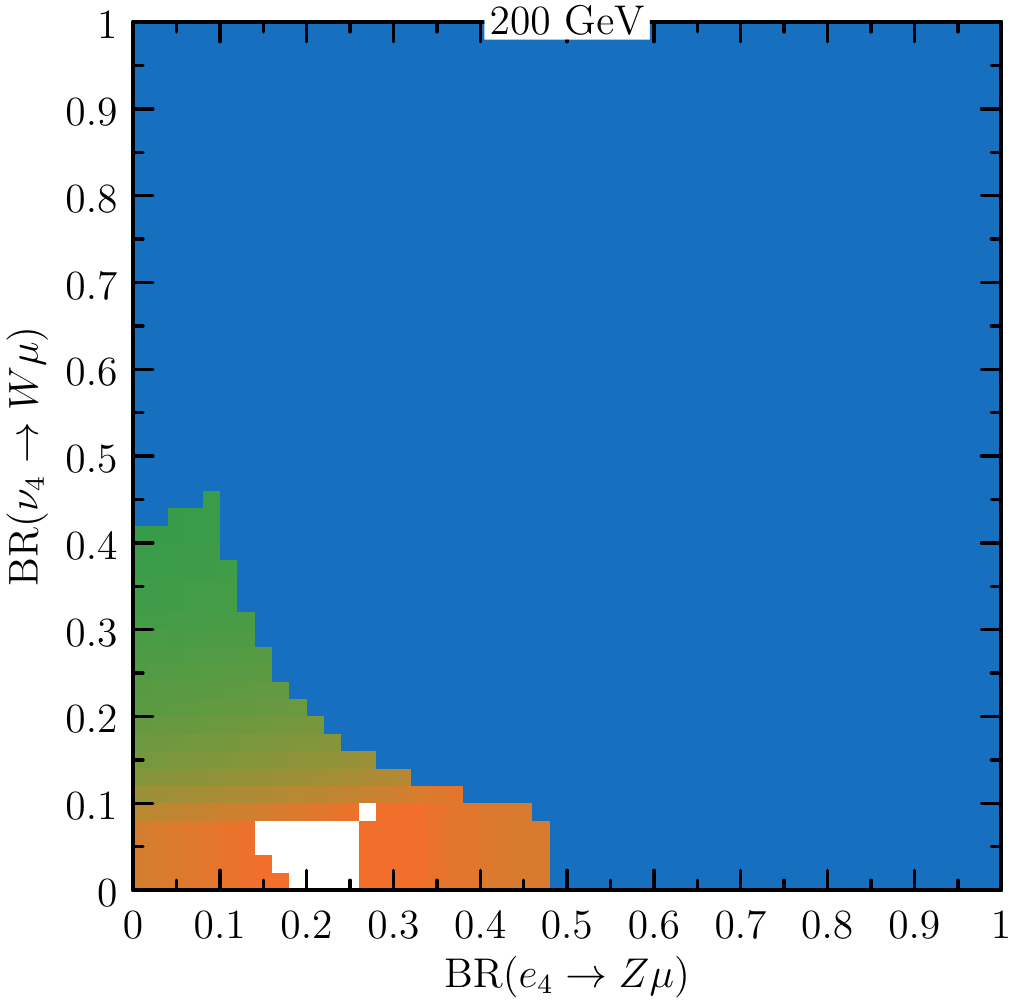}\\
\includegraphics[width=.45\linewidth]{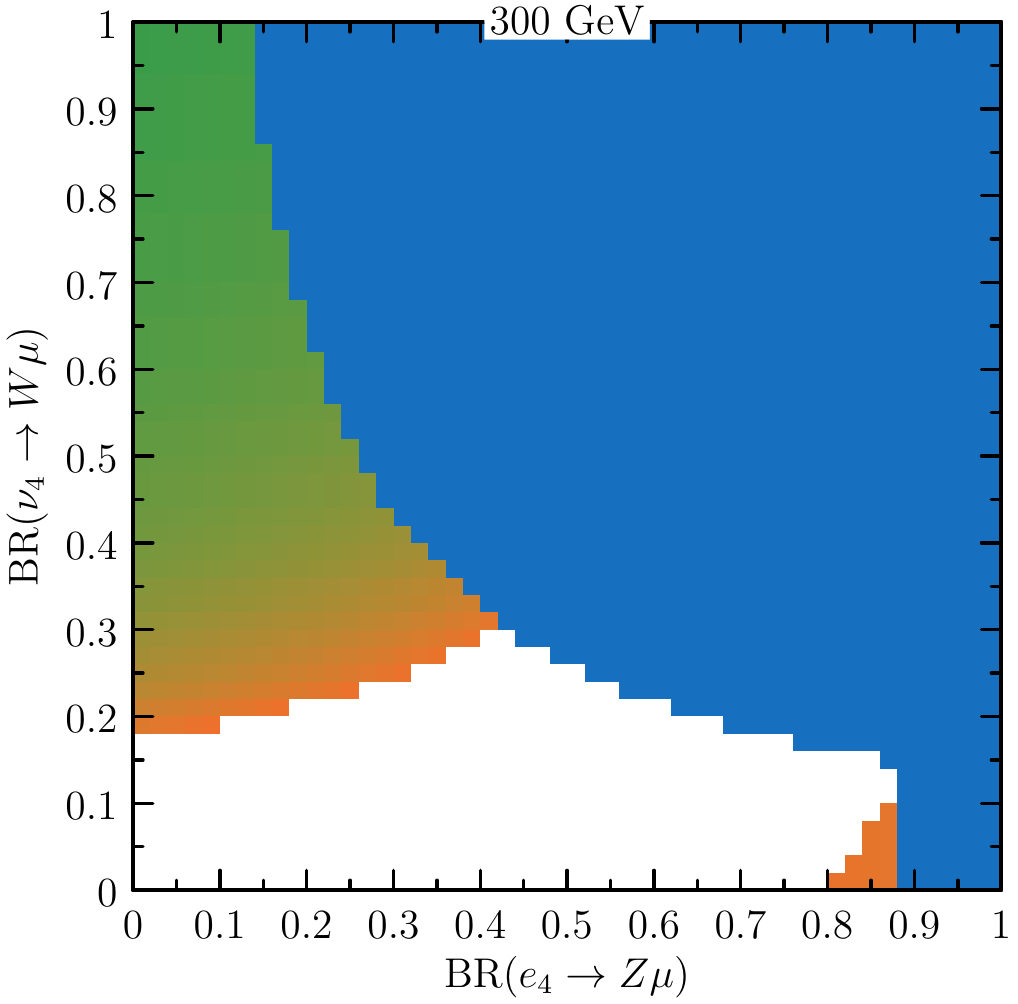}\hfill
\includegraphics[width=.45\linewidth]{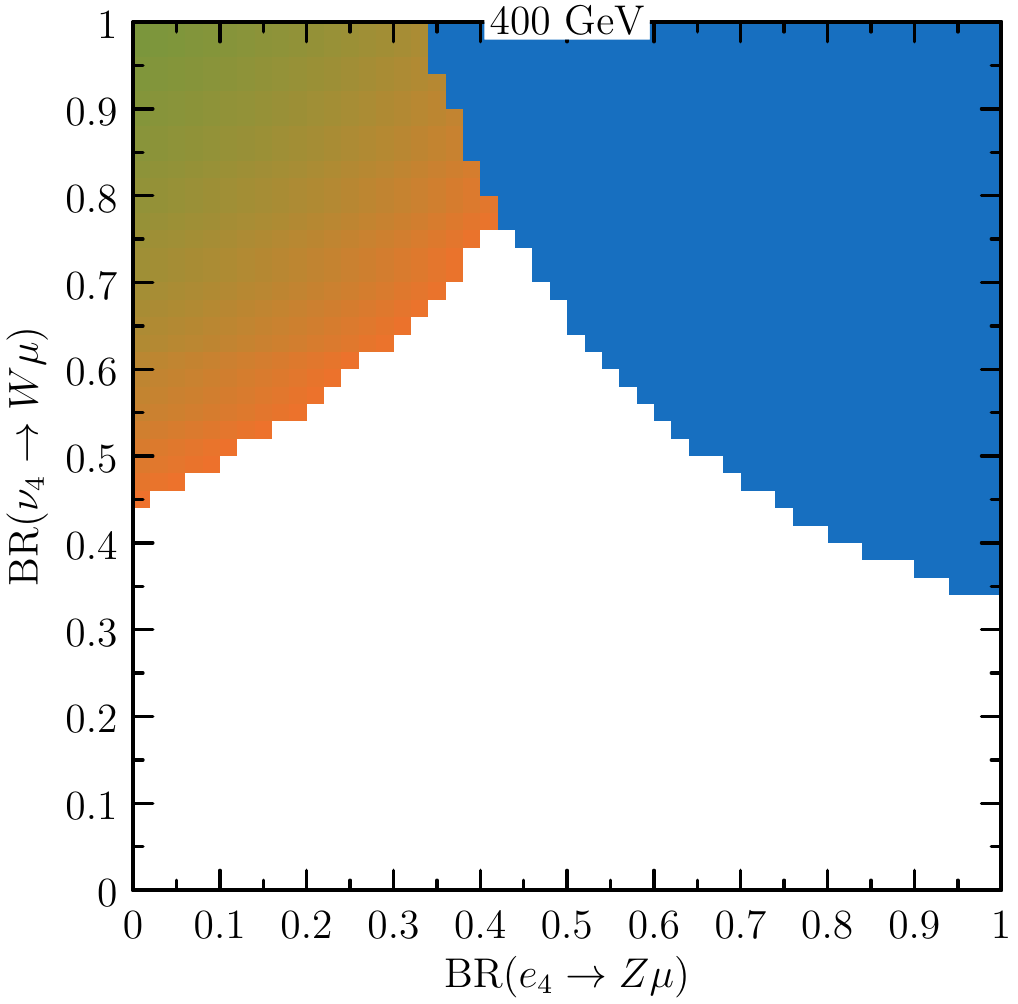}\\
\raisebox{-0.5\height}{\includegraphics[width=.45\linewidth]{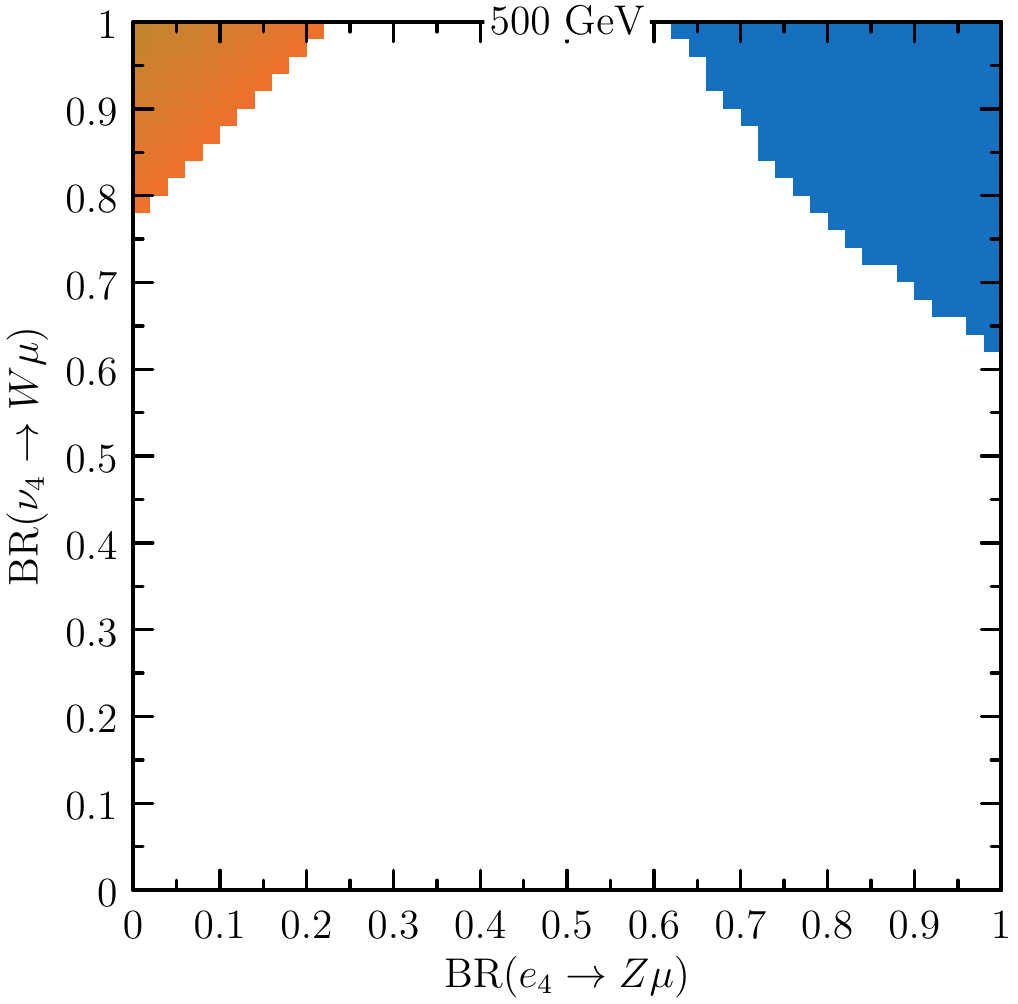}}\hfill
\raisebox{-0.5\height}{\includegraphics[width=.45\linewidth]{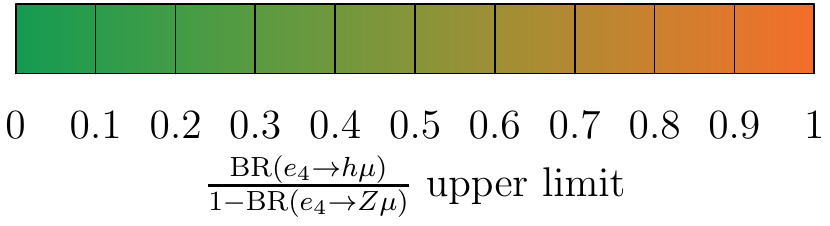}}
\caption{The general doublet case for different masses. Blue points are ruled out for all values of the other two independent branching ratios; white points are allowed for all values of the other two independent branching ratios; other points are shown in orange/green. The orange/green gradient shows the upper limit on the a third branching ratio divided by its maximum possible value
$\mathrm{BR}(e_4 \rightarrow h\mu)/(1-\mathrm{BR}(e_4 \rightarrow Z\mu))$.
\label{fig:da}}
\end{figure}

Above the Higgs threshold we have a four-dimensional parameter space and the results are more difficult to represent. There are six independent square projections through the hypercube. For 200 GeV all six are shown in figure~\ref{fig:da200}. Again the allowed space is in white and the ruled out space is in blue, but there is also now an orange region that contains both allowed and ruled out points in the four-dimensional parameter space. In the orange region contours can be plotted indicating the upper or lower limit on a third of the four independent branching ratios. In figure~\ref{fig:cont200} we show the upper (and, where relevant, lower) limits on a third independent branching ratio for the top two plots of figure~\ref{fig:da200}. These limits are indicated by an orange/purple colour gradient. In figure~\ref{fig:da} just the $[\rr{BR}(e_4\rightarrow Z\mu),\rr{BR}(\nu_4\rightarrow W\mu)]$ projection is shown for a range of masses, since these are the most directly constrained branching ratios of $e_4$ and $\nu_4$. In figure~\ref{fig:da} we choose to show additionally the upper limits on $\rr{BR}(e_4\rightarrow h\mu)$ since this is also quite well (directly) constrained and these limits are indicated by an orange/green colour gradient. The gradient only shows limits on this third branching ratio that are stronger than the trivial limit.

\section{Discussion and Conclusions}
\label{sec:conclusions}

In this paper we have derived some generally applicable limits on various heavy vectorlike lepton pair production and decay processes. We have achieved this by implementing the ATLAS analysis~\cite{TheATLAScollaboration:2013cia} and applying it to various processes generated within a general model. In the ATLAS paper single lepton (and single hadronic tau) fiducial efficiencies and particle level selection requirements are provided. Knowing these efficiencies or running a detector simulation that is good enough to reproduce them is very important when deriving limits from these multi-lepton analyses. These kinds of papers from ATLAS (and the {\tt Rivet}~\cite{Buckley:2010ar} implementations of the analyses) are therefore very useful for phenomenologists.

From our results we have also analysed the implied limits on branching ratios in various scenarios. We find that at this time, with the presently collected data, some combinations of branching ratios are poorly constrained, whereas some are constrained up to masses of about 300~GeV; the products of branching ratios $\rr{BR}(e_4 \rightarrow Z\mu)\rr{BR}(\nu_4 \rightarrow W \mu)$ and $\rr{BR}(e_4 \rightarrow h\mu)\rr{BR}(\nu_4 \rightarrow W \mu)$ are constrained up to masses of more than 500~GeV. We find that this implies that in some scenarios the parameter space is very well constrained at low masses, whereas in others with more freedom it is more open. In the doublet case with $\rr{BR}(\nu_4\rightarrow W\mu) = 1$ strong bounds are set. All masses below about 300~GeV are ruled out and there are strong limits constraining the branching ratios beyond 500~GeV. Alternatively, even if this condition is relaxed, below the Higgs threshold still almost all of the parameter space (of independent branching ratios) is ruled out. Nevertheless, it is interesting to point out that even assuming the maximal production cross-section, which coincides with the doublet case, the new lepton can still be as light as the LEP-II limit allows---105 GeV. We find that the sensitivity below the Higgs threshold could be improved even, with current data, by implementing a five-lepton cut. Results from such a cut can rule out the entire parameter space near 105 GeV and dramatically reduce the parameter space up to the Higgs threshold. Our results also indicate that $b$-tag cuts could be useful for processes involving at least one Higgs boson. Combining a five-lepton cut and $b$-tag cuts with the $p_T$ cuts in future analyses could be useful.

In conclusion, we stress that the limits we obtain are applicable to any model involving new pairs of neutral or charged particles allowed to decay to a vector or Higgs boson plus a charged lepton or neutrino (unless the kinematic structure of the dominant production mechanism differs significantly from production via $s$-channel SM bosons) and are already useful for constraining these models. Moreover the effectiveness of this kind of analysis will increase significantly after the next run of the LHC. Apart from the five-lepton cut that we discuss, another way in which the multi-lepton analysis could be tailored more specifically to heavy lepton searches involves $Z$ boson reconstruction. The decision in the ATLAS analysis~\cite{TheATLAScollaboration:2013cia} to only use the three leptons that define the event (rather than all reconstructed leptons) to attempt to reconstruct leptonically decaying $Z$ bosons has a significant effect on the categorization of our events, whereas it probably has little effect on the categorization of the background---many events are placed into the off-$Z$ categories that would otherwise be placed into the on-$Z$ categories. This effect is sometimes helpful and sometimes not. Perhaps more useful would be to use such reconstructed $Z$ bosons and SM leptons to reconstruct heavy lepton candidates and look for excesses in the invariant mass distribution. The ATLAS paper~\cite{ATLAS:2013hma}, analysing only 5.8~fb$^{-1}$ at 8~TeV, proceeds along these lines, searching for a signal in the reconstructed-leptonically-decaying-$Z$-boson-plus-charged-lepton invariant mass distribution. We encourage future searches of this kind as well as the general multi-lepton searches in order to constrain the possibility of (any kind of) heavy leptons. 

\acknowledgments{E.L. would like to thank Frank Siegert, Steffen Schumann and Clair Duhr for invaluable help regarding Sherpa and Rivet. R.D. and E.L. also thank the Mainz Institute for Theoretical Physics (MITP) for their hospitality and partial support during the completion of this work. S.S. thanks the Deutsches Elektronen-Synchrotron (DESY, Hamburg) for the support of visit during the process of this work. This work was supported in part by the Department of Energy under grant number DE-FG02-13ER42002.}

\end{document}